\documentclass[aps,prd,reprint,preprintnumbers,superscriptaddress,nofootinbib,amsmath,amssymb,longbibliography]{revtex4-2}
\usepackage[utf8]{inputenc}
\usepackage[dvipsnames]{xcolor}

\usepackage{tikz}
\usetikzlibrary{arrows.meta,positioning,bending}
\usepackage[export]{adjustbox}
\usepackage{mathtools,resizegather,amsmath,amsfonts,amssymb,mathrsfs,amsthm,amsbsy,multirow,dcolumn,makecell,scalerel,verbatim,braket,slashed, graphicx, physics,bm, hyperref, upgreek, xfrac}

\usepackage{amsthm}
\usepackage{orcidlink}
\newtheorem*{hypothesis}{Hypothesis (approximate kernel universality)}

\newcommand{\eref}[1]{Eq.~\eqref{#1}}

\def\be{\begin{equation}}
\def\ee{\end{equation}}

\begin{document}

\preprint{IFT-UAM/CSIC-26-107}

\title{Analytic Spread Complexity from Level Statistics: From Chaos to Integrability}



\author{Pallab Basu\,\orcidlink{0000-0003-0006-7240}}
\email{pallab.basu@wits.ac.za}
\affiliation{Mandelstam Institute for Theoretical Physics, School of Physics,\\ University of the Witwatersrand, Johannesburg, WITS 2050, South Africa.}
\author{Suman Das\,\orcidlink{0000-0002-0053-3187}}
\email{suman.das@ift.csic.es}
\affiliation{Instituto de F\'{i}sica Te\'{o}rica UAM/CSIC, Calle Nicol\'{a}s Cabrera 13-15, Madrid, E-28049, Spain.}
\author{Bigboy Madlala\,\orcidlink{0009-0006-2466-6781}}
\email{548440@students.wits.ac.za}
\affiliation{Mandelstam Institute for Theoretical Physics, School of Physics,\\ University of the Witwatersrand, Johannesburg, WITS 2050, South Africa.}

\begin{abstract}
Spread complexity has emerged as a useful probe of quantum chaos, yet the microscopic spectral origin of its characteristic finite-time peak remains incompletely understood. We develop an analytic framework that relates spread complexity directly to local spectral statistics. Starting from an
energy-space representation of the Krylov kernel, we show that the kernel is approximately banded, leading to a rapidly convergent diagonal expansion dominated by nearby levels in the ordered spectrum. Motivated by this structure, we propose an approximate kernel-universality hypothesis: after unfolding, the Krylov kernel is well approximated by that of a uniform lattice. Combining this universal kernel with local spectral statistics yields a simple analytic expression for spread complexity in terms of the Fourier transforms of the $k$th nearest-neighbour spacing distributions. In particular, at leading order, the finite-time peak is controlled by the Fourier transform of
the nearest-neighbour spacing distribution. The resulting framework describes both chaotic random-matrix ensembles and the integrable Poisson limit, identifies the spectral origin of the complexity peak and its late-time behaviour, and provides a direct connection between Krylov dynamics and spectral statistics.
\end{abstract}

\maketitle
\flushbottom

\section{Introduction and summary}

Characterizing the dynamics of isolated quantum many-body systems is a central problem in quantum statistical mechanics and quantum chaos. In recent years, quantum complexity has emerged as a useful dynamical probe that complements more traditional diagnostics such as out-of-time-order correlators~\cite{Shenker:2013pqa, Maldacena:2015waa}, spectral statistics \cite{PhysRevLett.52.1, Dyson:1963:STE4, PhysRevB.75.155111, PhysRevLett.110.084101, Mehta2004}, and the spectral form factor\cite{Dyson:1963:STE4, Cotler:2016fpe}. Among the various notions of complexity that have been proposed, Krylov complexity \cite{Parker:2018yvk, Balasubramanian:2022tpr} has attracted considerable attention because it admits a natural dynamical construction through the Lanczos algorithm~\cite{Lanczos1950} and can be efficiently computed in generic quantum systems~\cite{Dymarsky:2021bjq, Avdoshkin:2022xuw, Hashimoto:2023swv, Camargo:2023eev, Pal:2023yik, Nandy:2023brt,Bhattacharya:2023yec,Bhattacharyya:2023grv,Nizami:2024ltk, Huh:2023jxt, Nandy:2024wwv, Bhattacharjee:2024yxj, Basu:2025ubf,Balasubramanian:2024ghv, Chatzis:2026ekd, Roychowdhury:2026vzq, Caputa:2024vrn, Imani:2025etp, Banerjee:2026kav, Baume:2026jyt, Chakraborty:2026ssu}. Owing to these properties, Krylov complexity has found applications in the study of operator growth,
information scrambling, thermalization, transport, and quantum chaos (for recent review see~\cite{Nandy:2024evd, Rabinovici:2025otw}).

In holography~\cite{Maldacena:1997re,Witten:1998qj,Aharony:1999ti}, quantum complexity has been related to bulk geometry through the complexity--volume, complexity--action and complexity--anything proposals~\cite{Stanford:2014jda, Couch:2016exn, Brown:2015bva, Belin:2021bga, Belin:2022xmt}. Krylov-based notions of complexity provide a complementary perspective. In double-scaled SYK, the Krylov basis of the infinite-temperature thermofield-double state admits, in the appropriate JT limit, a bulk interpretation in terms of length states, relating spread complexity to the growth of a two-sided wormhole~\cite{Lin:2022rbf,Rabinovici:2023yex,Aguilar-Gutierrez:2025pqp,Fu:2025kkh, Alfinito:2026cky}. Related developments connect its growth rate to the proper radial momentum of a bulk excitation and its late-time saturation to that of the Einstein--Rosen bridge~\cite{Caputa:2024sux,Balasubramanian:2024lqk, Alfinito:2026vah}. Since universal spectral correlations govern late-time quantum-black-hole dynamics~\cite{Cotler:2016fpe}, their imprint on spread complexity may help clarify the microscopic origin of these bulk-geometric phenomena.

Two closely related notions of Krylov complexity have been developed. Operator Krylov complexity~\cite{Parker:2018yvk} quantifies the growth of operators under repeated commutation with the Hamiltonian and has proved to be a useful diagnostic of operator spreading and quantum chaos. A complementary notion, usually referred to as state or spread Krylov complexity, was introduced in~\cite{Balasubramanian:2022tpr}, where complexity is defined through the spreading of a quantum state under successive applications of the Hamiltonian. In this work, we focus exclusively on spread Krylov complexity and refer to it simply as spread complexity throughout.

One of the most intriguing observations concerning spread complexity is its characteristic time dependence. For chaotic systems, the complexity typically exhibits a growth--peak--plateau structure. Remarkably, the height of the intermediate peak is not universal. Random matrix studies have shown that it depends sensitively on the underlying spectral statistics and varies across different symmetry classes \cite{Erdmenger:2023wjg}. More generally, for ensembles interpolating between Poisson and Wigner--Dyson statistics, the peak height decreases continuously as level repulsion is weakened and eventually disappears in the
integrable limit \cite{Scialchi:2023bmw, Camargo:2024deu, Huh:2024ytz, Baggioli:2024wbz}. These observations suggest that the peak contains valuable information about the underlying spectral correlations and may serve as a quantitative probe of quantum chaos.

A dynamical explanation of the complexity peak was proposed in ref.~\cite{Erdmenger:2023wjg} (also see \cite{Alishahiha:2024vbf, Baggioli:2024wbz}) by analysing the transition probabilities between the initial state and individual Krylov-basis states. For chaotic spectra, these probabilities exhibit a characteristic rise--slope--ramp--plateau structure\footnote{This linear ramp is reminiscent of the universal ramp of the spectral form factor, both being associated with chaotic spectral correlations. However, in \cite{Das:2023yfj,Basu:2025zkp} it has been shown that a linear ramp can also exist in the $\beta=0$ spectral form factor for deterministic spectra such as $E_n=\{\log n\}$. As a result, the spread complexity can also exhibit a peak, as shown in
\cite{Jeong:2024jjn,Begines:2026fnx, Jeong:2025jyx}. }, which was argued to provide the dynamical origin of the temporary overshoot of the spread complexity above its late-time plateau. In contrast, the ramp is absent for uncorrelated spectra, resulting in a monotonic approach to equilibrium. While this provides an appealing dynamical interpretation of the peak, it does not relate its height or its
symmetry-class dependence directly to the underlying spectral statistics. In particular, it is still unclear which spectral correlations control the peak, why its height varies between different random-matrix ensembles, and how spread complexity can be related quantitatively to standard measures of spectral statistics.

In this work we address these questions by reformulating spread complexity directly in the energy eigenbasis. Instead of viewing the dynamics solely as the spreading of a state in Krylov space, we derive an exact representation in which the complexity is expressed as a double sum over energy eigenstates \eref{eq:KC_energy}. The result naturally separates into two ingredients: a structural kernel that
depends only on the Krylov basis, and oscillatory phase factors determined by the energy spectrum \eref{eq:C0_Ck_definition}. This decomposition provides a direct bridge between Krylov dynamics and spectral statistics and forms the basis of our subsequent analysis.

A central result of this work is that the Krylov kernel possesses an unexpected quasi-local, banded structure in the energy basis (see Fig~\ref{fig:K_ij_matrixplot}). Although the complexity operator is nonlocal in Hilbert space, its energy-space kernel is concentrated near the diagonal, reflecting the fact that the two-point function of the relevant Krylov basis---orthogonal polynomials in the energy basis---decays with separation. In addition, the kernel exhibits an approximately semicircular profile (see Fig~ \ref{fig:Kernel_diagonal_reverse}), which follows from the finite support of these polynomials in the bulk of the spectrum. This observation leads to a natural diagonal expansion in which each term is associated with energy levels separated by a fixed number of spacings. We find that the corresponding diagonal weights decay approximately as $k^{-2}$, implying that only the first few neighboring levels contribute significantly to the dynamics (see Fig~\ref{fig:Krylov_comparison}). This provides a microscopic understanding of why short-range spectral correlations dominate the peak structure.

The quasi-local structure of the kernel further motivates an approximate kernel-universality hypothesis (Sec.~\ref{hyp:kernel_universality}). After unfolding the spectrum, we find that the orthogonal-polynomial basis associated with generic random-matrix spectra is related to that of a uniform lattice by an approximately local transformation. Consequently, the unfolded Krylov kernel is  approximated well by the analytically tractable kernel of the uniform lattice. This is reminiscent of a broader pattern in physics, in which simplified, solvable models capture significant physics, from the Ising and Curie--Weiss models of magnetism to the Anderson--Yuval--Hamann treatment of the Kondo problem \cite{Onsager1944,Weiss1907,GinzburgLandau1950,Anderson1970,AndersonYuvalHamann1970}.

Combining this universal kernel with unfolded spectral statistics yields a simple master formula expressing spread complexity in terms of the Fourier
transforms of the $k$th nearest-neighbour spacing distributions\footnote{In \cite{FarajiAstaneh:2025rlc}, the connection between complexity and the level-spacing distribution was discussed for a two-level system.}.
\begin{equation}
\frac{\langle C_{\rm unfold}(t) \rangle}{D}
\simeq
\frac12 \left(1
-
2\sum_{k=1}^{\infty}
\frac{\Phi_k(t)}
{4k^2-1} \right).
\end{equation}
Within this framework, the universal kernel determines the relative importance of different neighbour separations, while all ensemble dependence enters through the local spectral statistics. Keeping only the leading contribution already reproduces the peak height of the GOE and GUE ensembles with remarkable accuracy (see Fig~ \ref{fig:complexity_GOE_GUE}). Furthermore, for Poisson statistics the full series can be resummed analytically, demonstrating that the complexity approaches its plateau monotonically without developing a finite-time peak. Our results therefore provide a microscopic framework that connects spread complexity directly to local spectral statistics and identifies short-range level correlations as the origin of the complexity peak.

The remainder of this paper is organized as follows. In section~\ref{sec:preliminaries}, we review spread complexity and derive its energy-space kernel representation and diagonal decomposition. In section~\ref{sec:banded_kernel}, we analyze the banded structure and semicircular
profile of the Krylov kernel. In section~\ref{sec:the_hypothesis}, we introduce and test the approximate kernel-universality hypothesis. Sections~\ref{sec:analytic_peak} and~\ref{sec:Poisson_no_peak} apply this framework to the Gaussian ensembles and Poisson statistics, respectively. In section~\ref{sec:syk}, we test the framework on the Sachdev--Ye--Kitaev model, an interacting many-body system rather than a spectral ensemble. We conclude in section~\ref{sec:discussions}. Further technical details are provided in appendices~\ref{app:uniform_kernel_sums}, \ref{app:folded_unfolded_time} and \ref{app:poisson_checks}.

\section{Spread complexity and its kernel representation}
\label{sec:preliminaries}

Let $H$ be a time-independent Hermitian Hamiltonian acting on a $D$-dimensional Hilbert space, and let $\ket{\psi_0}$ be a normalized initial state. Throughout this work, we set $\hbar=1$. The time-evolved state is
\begin{align}
    \ket{\psi(t)}
    &=
    e^{-iHt}\ket{\psi_0}
    =
    \sum_{\ell=0}^{\infty}
    \frac{(-it)^\ell}{\ell!}
    H^\ell\ket{\psi_0}.
\end{align}
The vectors $\{H^\ell\ket{\psi_0}\}$ do not, in general, form an orthonormal basis. Applying a Gram--Schmidt-type procedure, namely the Lanczos algorithm, one obtains an orthonormal Krylov basis $\{\ket{K_n}\}_{n=0}^{K-1}$ satisfying
\begin{equation}
\begin{gathered}
H\ket{K_n}
    =
    b_{n+1}\ket{K_{n+1}}
    +
    a_n\ket{K_n}
    +
    b_n\ket{K_{n-1}},
    \\
    \ket{K_{-1}}=0,
    \qquad
    b_0=0,
    \qquad
    \ket{K_0}=\ket{\psi_0}.
\end{gathered}
\label{eq:Lanczos_recursion}
\end{equation}
For a finite Krylov chain, the recursion terminates at $b_K=0$. The coefficients $a_n$ and $b_n$ are referred to as the Lanczos coefficients, while the space spanned by $\{\ket{K_n}\}_{n=0}^{K-1}$ is called the \emph{Krylov subspace}.

The time-evolved state can be expanded in the Krylov basis as
\begin{equation}
    \ket{\psi(t)}
    =
    \sum_{n=0}^{K-1}
    \phi_n(t)\ket{K_n},
    \qquad
    \phi_n(t)
    =
    \braket{K_n}{\psi(t)}.
\end{equation}
The coefficients $\phi_n(t)$ are the probability amplitudes of the evolved state in the Krylov basis.

Each Krylov vector can equivalently be expressed as a polynomial of the Hamiltonian acting on the initial state~\cite{Muck:2022xfc, Balasubramanian:2026azk},
\begin{equation}
    \ket{K_n}
    =
    p_n(H)\ket{\psi_0},
    \label{eq:defn_poly}
\end{equation}
where $p_n(E)$ is a polynomial of degree $n$. The corresponding polynomial family satisfies
\begin{equation}
\begin{gathered}
E\,p_n(E)
    =
    b_{n+1}p_{n+1}(E)
    +
    a_n p_n(E)
    +
    b_n p_{n-1}(E),
    \\
    p_{-1}(E)=0,
    \qquad
    p_0(E)=1.
\end{gathered}
\label{eq:polynomial_recursion}
\end{equation}
Since the Krylov basis is orthonormal, these polynomials are orthonormal with respect to the spectral measure induced by the initial state. To make this explicit, we expand the initial state in the orthonormal energy basis
\begin{equation}
    \ket{\psi_0}
    =
    \sum_{i=0}^{D-1}
    c_i\ket{E_i}, \qquad
    \sum_{i=0}^{D-1}|c_i|^2=1.
    \label{eq:initial_state_energy_basis}
\end{equation}
Using eq.~\eqref{eq:defn_poly}, one finds
\begin{equation}
    \ket{K_n}
    =
    \sum_{i=0}^{D-1}
    c_i \, p_n(E_i)\ket{E_i}.
\end{equation}
The orthonormality condition $\braket{K_m}{K_n}=\delta_{mn}$ therefore gives
\begin{equation}
    \sum_{i=0}^{D-1}
    |c_i|^2
    p_m(E_i)^*p_n(E_i)
    =
    \delta_{mn}.
    \label{eq:polynomial_orthogonality_general}
\end{equation}
For a Hermitian Hamiltonian, the Lanczos coefficients may be chosen real, and hence the polynomials $p_n(E)$ may also be chosen real. With this convention, eq.~\eqref{eq:polynomial_orthogonality_general} reduces to
\begin{equation}
    \sum_{i=0}^{D-1}
    |c_i|^2
    p_m(E_i)p_n(E_i)
    =
    \delta_{mn}.
    \label{eq:polynomial_orthogonality}
\end{equation}
Thus, the Lanczos polynomials form an orthonormal family with respect to the discrete spectral measure
\begin{equation}
    d\mu(E)
    =
    \sum_{i=0}^{D-1}
    |c_i|^2
    \delta(E-E_i)\,dE
\end{equation}
induced by the initial state.

Using the above energy decomposition, the Krylov amplitudes become
\begin{equation}
    \phi_n(t)
    =
    \sum_{i=0}^{D-1}
    |c_i|^2
    p_n(E_i)
    e^{-iE_it},
    \qquad
    n=0,\ldots,K-1.
    \label{eq:com_op}
\end{equation}
The Krylov position operator is defined by
\begin{equation}
    \widehat{\mathcal C}
    =
    \sum_{n=0}^{K-1}
    n\,\ket{K_n}\bra{K_n}.
    \label{eq:Krylov_position_operator}
\end{equation}
The spread complexity is the expectation value of this operator in the time-evolved state,
\begin{equation}
    C(t)
    =
    \bra{\psi(t)}
    \widehat{\mathcal C}
    \ket{\psi(t)}
    =
    \sum_{n=0}^{K-1}
    n\,|\phi_n(t)|^2.
    \label{eq:Krylov_complexity_definition}
\end{equation}
Substituting the energy-space expression for $\phi_n(t)$ into eq.~\eqref{eq:Krylov_complexity_definition}, one obtains
\begin{align}
    C(t)
    &=
    \sum_{n=0}^{K-1}
    n
    \sum_{i,j=0}^{D-1}
    |c_i|^2|c_j|^2
    p_n(E_i)p_n(E_j)
    e^{-i(E_i-E_j)t}
    \nonumber\\
    &=
    \sum_{i,j=0}^{D-1}
    |c_i|^2|c_j|^2
    e^{-i(E_i-E_j)t}
    \sum_{n=0}^{K-1}
    n\,p_n(E_i)p_n(E_j).
    \label{eq:complexity_before_kernel}
\end{align}
We now define the normalized Krylov kernel by
\begin{equation}
    \mathcal K(i,j)
    =
    \frac{1}{D^2}
    \sum_{n=0}^{K-1}
    n\,p_n(E_i)p_n(E_j)
    \label{eq:kernel_ij}.
\end{equation}
In terms of the normalized kernel, the spread complexity for a general initial state assumes the exact form
\begin{equation}
    C(t)
    =
    D^2
    \sum_{i,j=0}^{D-1}
    |c_i|^2|c_j|^2
    \mathcal K(i,j)
    e^{-i(E_i-E_j)t}.
    \label{eq:KC_energy_general}
\end{equation}
The kernel $\mathcal K(i,j)$ depends on the Lanczos polynomials, and therefore on both the underlying energy spectrum and the spectral measure induced by the initial state. Equation~\eqref{eq:KC_energy_general} separates the complexity into a structural kernel and oscillatory phases determined by the energy differences.

For simplicity, in the remainder of this work we consider the infinite-temperature thermofield-double state. In this case, the initial-state coefficients are uniformly distributed,
\begin{equation}
    |c_i|^2
    =
    \frac{1}{D},
    \qquad
    i=0,\ldots,D-1.
    \label{eq:uniform_TFD_coefficients}
\end{equation}
The orthonormality relation for the Lanczos polynomials consequently becomes
\begin{equation}
    \frac{1}{D}
    \sum_{i=0}^{D-1}
    p_m(E_i)p_n(E_i)
    =
    \delta_{mn}.
    \label{eq:polynomial_orthogonality_TFD}
\end{equation}
Substituting eq.~\eqref{eq:uniform_TFD_coefficients} into eq.~\eqref{eq:KC_energy_general}, the factors of $D$ cancel and the complexity takes the particularly simple form
\begin{equation}
    C(t)
    =
    \sum_{i,j=0}^{D-1}
    \mathcal K(i,j)
    e^{-i(E_i-E_j)t}.
    \label{eq:KC_energy}
\end{equation}

Equation~\eqref{eq:KC_energy} shows that the complexity is determined by the competition between the oscillatory phase associated with the energy difference $E_i-E_j$ and the kernel $\mathcal K(i,j)$, which specifies the weight carried by each pair of energy eigenstates. This representation provides a natural framework for analysing the contributions from different energy-level separations.

Assuming that the energy eigenvalues are arranged in increasing order, the double sum in eq.~\eqref{eq:KC_energy} may be reorganized according to the distance from the principal diagonal of the kernel matrix. Since the kernel is real and symmetric, the complexity may be decomposed as \footnote{ See \cite{Martinez-Azcona:2023umk} for a similar analysis of the SFF.}
\begin{equation}
    C(t)
    =
    C_0
    +
    \sum_{k=1}^{D-1}
    C_k(t),
    \label{eq:complexity_sum1}
\end{equation}
where the diagonal  and the $k$-th off-diagonal contributions are given by
\begin{equation}
\begin{gathered}
C_0 = \sum_{i=0}^{D-1} \mathcal K(i,i),
    \\
    C_k(t) = 2 \sum_{i=0}^{D-k-1} \mathcal K(i,i+k) \cos\!\left[(E_{i+k}-E_i)t\right],
\end{gathered}
\label{eq:C0_Ck_definition}
\end{equation}
This decomposition separates the exact spread complexity into contributions from different energy-level separations, with $C_k(t)$ determined solely by pairs of levels separated by $k$ positions in the ordered spectrum.

\begin{figure}
    \centering
    \includegraphics[width=\linewidth]{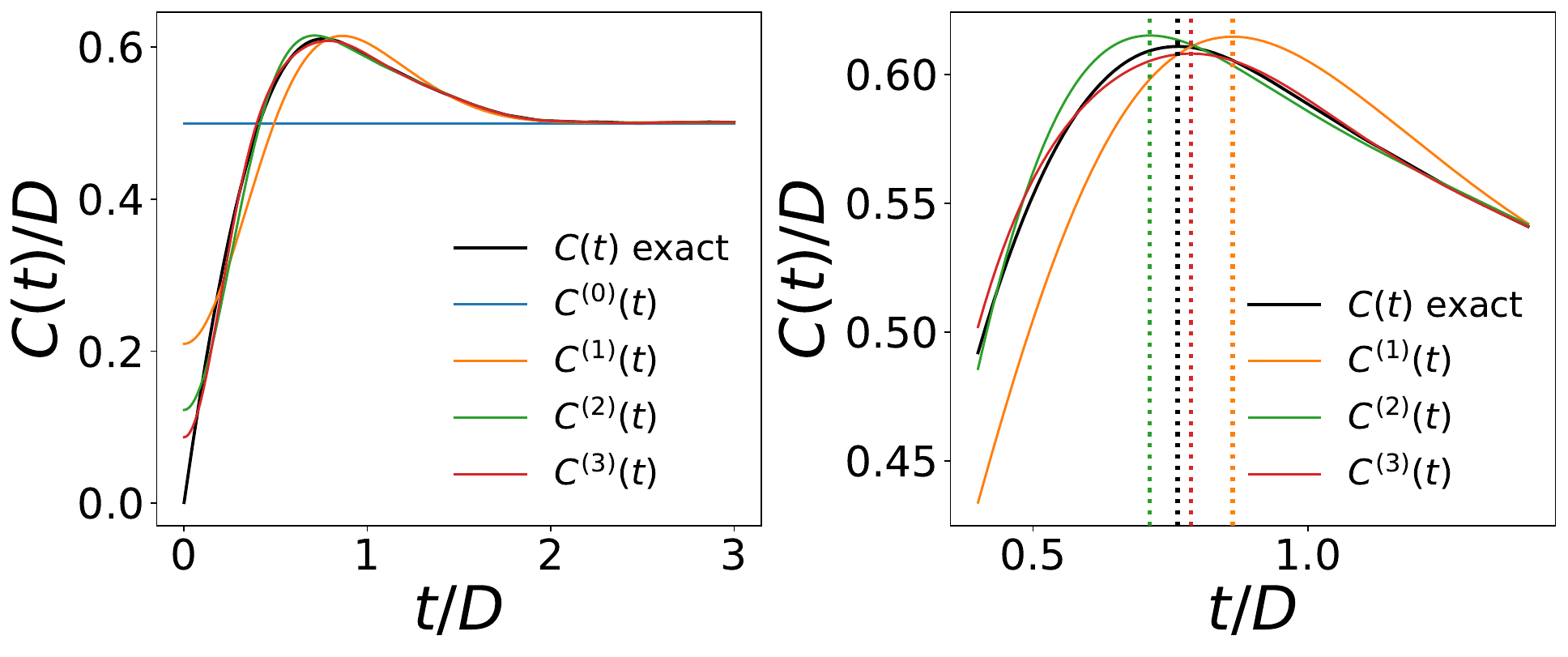}
    \caption{Comparison of the exact spread complexity $C(t)$ with the successive approximations $C^{(0)}(t)=C_0$, $C^{(1)}(t)$, $C^{(2)}(t)$, and $C^{(3)}(t)$, computed from raw (not unfolded) GUE spectra of dimension $D=1000$ and averaged over 8 realizations. The left panel shows the full time evolution, while the right panel shows an enlarged view near the maximum. The vertical dotted lines indicate the corresponding peak positions. Retaining only the first few diagonals of the Krylov kernel accurately reproduces the exact complexity.}
    \label{fig:Krylov_comparison}
\end{figure}

A crucial observation of this paper is that the expansion in Eq.~\eqref{eq:complexity_sum1} converges rapidly to full complexity, even for a single realization of the spectrum, without requiring any ensemble averaging. Consequently, the late-time behavior of $C(t)$ is accurately captured by retaining only the first few terms in the expansion. Fig.~\ref{fig:Krylov_comparison} compares the exact spread complexity with the successive approximations $C^{(m)}(t)$\footnote{Where\begin{equation} C^{(m)}(t) = C_0 + \sum_{k=1}^{m} C_k(t), \end{equation}denotes the complexity truncated to the first $m+1$ kernel diagonals.}. It is well known that at late times the spread complexity approaches its diagonal ensemble value $C_0$~\cite{Baggioli:2024wbz, Alishahiha:2024vbf}, despite the fact that complexity itself is not an operator for which ETH is expected to apply. The rapid convergence at small $m$ indicates that Krylov dynamics is primarily governed by nearby energy-level separations, with contributions from increasingly distant levels progressively suppressed.

Table~\ref{tab:peak_comparison} compares the peak position and height obtained from the exact result with those obtained from the truncated approximations. The percentage errors decrease as $m$ increases, confirming that only a few kernel diagonals are sufficient to capture the dominant features of the exact dynamics, including the peak.
\begin{table}[h]
\centering
\begin{tabular}{lcc}
\hline
Method & $t_{\mathrm{peak}}/D$ & $C(t_{\mathrm{peak}})/D$ \\
\hline
Exact    & 0.76287 & 0.61094 \\
Approx 1 & 0.86320 (+13.15\%) & 0.61472 (+0.62\%) \\
Approx 2 & 0.71259 (-6.59\%)  & 0.61518 (+0.69\%) \\
Approx 3 & 0.78717 (+3.19\%)  & 0.60815 (-0.46\%) \\
Approx 4 & 0.75857 (-0.56\%)  & 0.61262 (+0.28\%) \\
Approx 5 & 0.75162 (-1.47\%)  & 0.61031 (-0.10\%) \\
Approx 6 & 0.77156 (+1.14\%)  & 0.61110 (+0.03\%) \\
\hline
\end{tabular}
\caption{Comparison of the peak position and peak height of the exact spread complexity with the successive approximations $C^{(m)}(t)$. The numbers in parentheses denote the percentage deviations relative to the exact result.}
\label{tab:peak_comparison}
\end{table}

These observations stem from an important property of the Krylov kernel $\mathcal K(i,j)$: not all kernel elements contribute equally to the complexity; instead, the dominant contributions arise from a narrow band around the principal diagonal, indicating that the kernel is approximately banded. Moreover, along the first few diagonals the kernel decays via a sequence of discrete jumps. To our knowledge, this decomposition has not appeared elsewhere. While the sine kernel is useful in certain situations (e.g., Eq.~31 of \cite{Alishahiha:2024vbf}), it obscures the fact that, unlike the SFF, the Krylov-complexity kernel can vary rapidly in the direction transverse to the principal diagonal; a formulation in terms of the level spacing, as adopted here, is therefore more appropriate. We investigate this banded structure in detail in the next section.

\section{Krylov kernel and its banded structure}
\label{sec:banded_kernel}

\begin{figure}[tbp]
    \centering
    \includegraphics[width=0.7\linewidth]{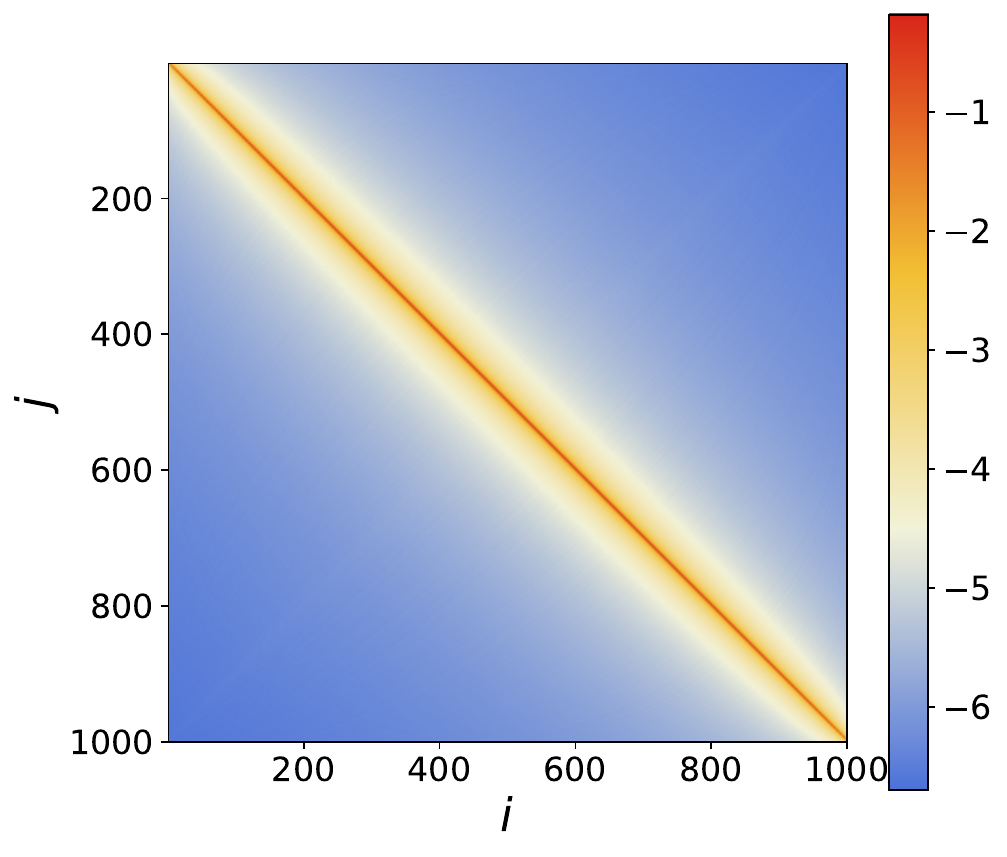}
    \caption{Heat map of the Krylov kernel $\mathcal{K}(i,j)$. The colour scale
    represents
    $\log_{10}\!\left(\langle |\mathcal{K}(i,j)|\rangle\right)$, where the
    ensemble average is taken over $50$ realizations. The logarithmic scale is
    used to enhance the visibility of the off-diagonal structure. The figure
    clearly illustrates the approximately banded nature of the kernel, with
    the magnitude of its matrix elements decreasing as one moves away from the
    principal diagonal. Although an ensemble average is shown here, the same
    qualitative behaviour is observed for individual realizations.}
    \label{fig:K_ij_matrixplot}
\end{figure}

In the previous section, we showed that the spread complexity can be decomposed into contributions arising from different diagonals of the kernel matrix. The usefulness of this decomposition depends on the structure of the kernel itself. In this section, we demonstrate that the Krylov kernel possesses an approximately banded structure. To illustrate this, Fig.~\ref{fig:K_ij_matrixplot} shows the heat map of the kernel $\mathcal{K}(i,j)$, from which it is evident that the magnitude of the matrix elements decreases rapidly as one moves away from the principal diagonal. 

To quantify the contribution from each off-diagonal, we define
\begin{equation}
\label{eq:W_k_defn}
W_k
=
\sum_{i=0}^{D-k-1}
\mathcal K(i,i+k),
\end{equation}
which measures the net integrated weight of the kernel along the $k$-th off-diagonal.  As $C(0)=0$, we have,
\begin{align}
\label{eq:sum_rule_Wk}
   W_{0}+2 \sum_{k=1}^{D-1} W_{k}=0
\end{align}
From the trace we obtain $W_0=\frac{1}{D}\Tr \widehat {\mathcal C} =\tfrac12(D-1)$. For $k\ge1$, the weights $W_k$ are often negative. In particular, $W_k$ is a $k$-shift autocorrelation of the orthogonal polynomials, with contributions weighted by the polynomial degree; since higher-degree polynomials oscillate rapidly, it is therefore not surprising that $W_1<0$.

\begin{figure}[t]
    \centering
    \includegraphics[width=1\linewidth]{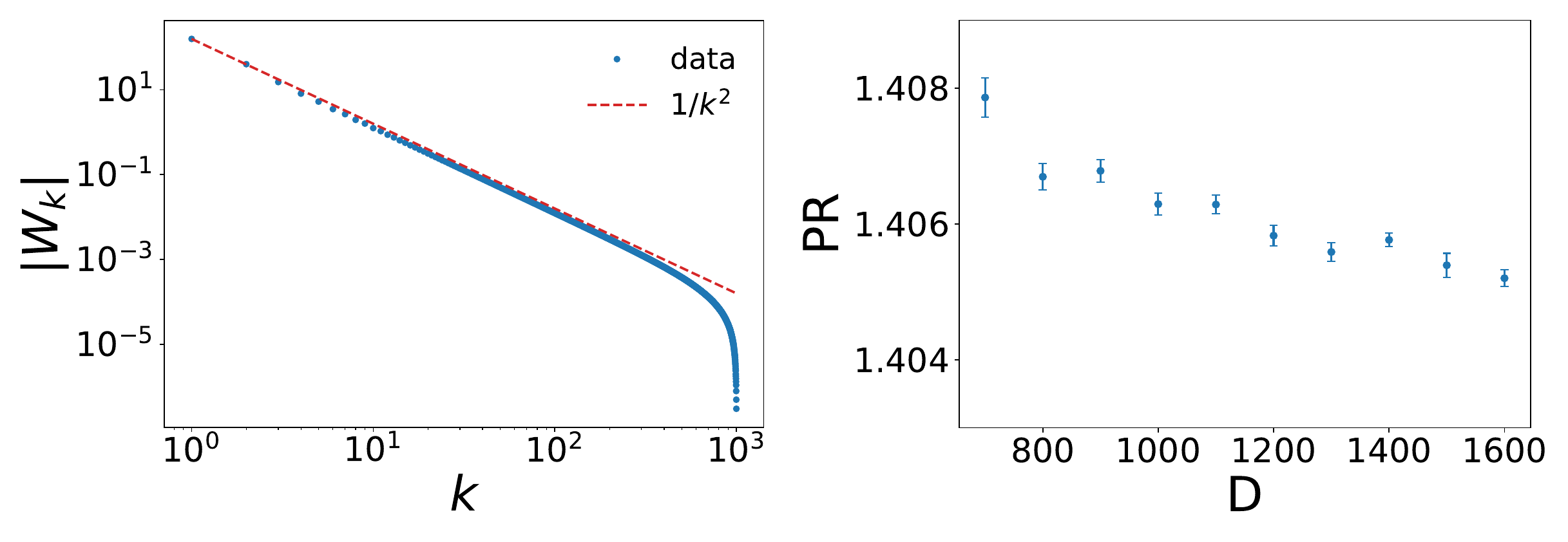}
    \caption{Left: the integrated off-diagonal weight $|W_k|$,
    eq.~\eqref{eq:W_k_defn}, versus $k$ on a log--log scale, showing an
    approximate $1/k^2$ decay and confirming the banded structure of the
    kernel. Right: the disorder-averaged participation ratio
    $\mathrm{PR}$ versus $D$.}
    \label{fig:Wk_vs_k}
\end{figure}

Over sufficiently large $k$, the corresponding polynomials become increasingly uncorrelated, so one expects the magnitude $|W_k|$ to decay with $k$. Fig.~\ref{fig:Wk_vs_k} shows $|W_k|$ as a function of the off-diagonal index $k$ on a log--log scale. The data exhibit an excellent $1/k^2$ decay over a broad range of $k$, indicating a rapid suppression of the net weight carried by distant kernel diagonals. This decay therefore provides a useful empirical measure of how quickly successive diagonals lose weight with energy-level separation, and it suggests that only the first few diagonals make a dominant contribution to the spread complexity.

\begin{figure}
    \centering
    \includegraphics[width=1\linewidth]{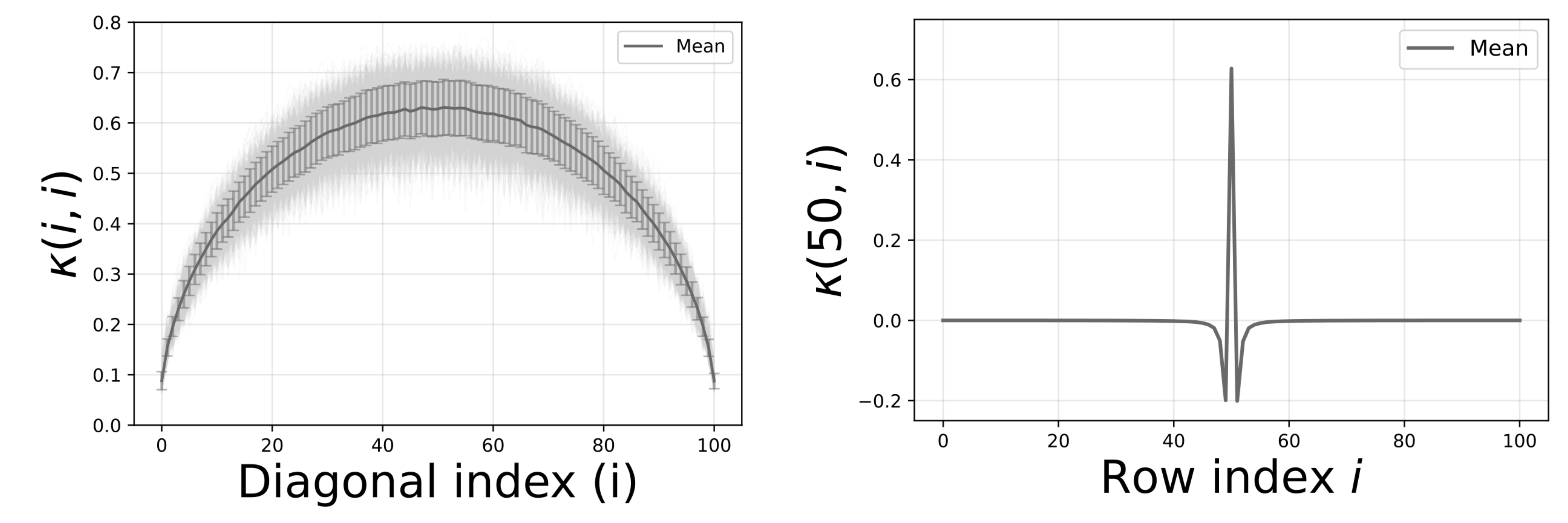}
    \caption{Diagonal and row-wise profiles of the kernel $\mathcal{K}(i,j)$ computed from a single raw GUE spectrum of dimension $D=101$. Left: the main diagonal $\mathcal{K}(i,i)$, showing an approximately semicircular envelope together with realization-specific fluctuations. Right: the middle row $\mathcal{K}\bigl((D-1)/2,i\bigr)$, which remains essentially zero except for a single sharp feature near $i\simeq D/2$, where the row crosses the main diagonal. This confirms that $\mathcal{K}$ is strongly peaked near $i=j$ and decays rapidly away from it.}
    \label{fig:Kernel_diagonal_reverse}
\end{figure}

An important consequence of Eq.~\eqref{eq:kernel_ij} is what we call \emph{semicircularity}: the relevant orthogonal polynomials typically have support concentrated near the middle of the spectral domain, so that for finite $D$ the kernel fluctuates around an approximately semicircular mean value,
\begin{align}
    \mathcal{K}(i,j) = \mathcal{K}_{\text{mean}}(i,j)\,\big(1+\delta_{\text{err}}(i,j)\big),
    \label{eq:kernel_mean_error}
\end{align}
where $\delta_{\text{err}}(i,j)$ is an error term expected to scale as $O(1/\sqrt{D})$; we leave a detailed analysis of this term for future work. Concretely, this means that along a fixed diagonal (or off-diagonal) the kernel is largest near the center of the domain and decays toward both ends, producing the approximately semicircular profile illustrated in Fig.~\ref{fig:Kernel_diagonal_reverse} (left) (also see \eref{eq:semi}).

The two directions in the $(i,j)$ plane behave quite differently. Along the direction parallel to the principal diagonal, the kernel varies smoothly, with fluctuations of order $O(1/D)$. In the transverse direction, by contrast, it can vary strongly and exhibit $O(1)$ discrete jumps, as illustrated by the row profile in Fig.~\ref{fig:Kernel_diagonal_reverse}(right). This anisotropic structure implies that different off-diagonal sectors contribute very differently to the spread complexity, which motivates the decomposition in Eq.~\eqref{eq:C0_Ck_definition}.

\section{Approximate kernel universality hypothesis}
\label{sec:the_hypothesis}

In the previous sections, we showed that the complexity is determined by the Krylov kernel $\mathcal K(i,j)$. In the following, we restrict ourselves to a nondegenerate spectrum for which the initial state has non-vanishing overlap with all $D$ energy eigenstates. The Krylov dimension is then maximal, 
\begin{equation}
K=D.
\label{eq:maximal_Krylov_dimension}
\end{equation}
For the infinite-temperature TFD state, the Krylov kernel associated with a spectrum $\{E_i\}_{i=0}^{D-1}$ is therefore
\begin{equation}
\mathcal K(i,j)
=
\frac{1}{D^2}
\sum_{n=0}^{D-1}
n\,p_n(E_i)p_n(E_j),
\label{eq:kernel_general_recalled}
\end{equation}
where $\{p_n(E)\}$ denotes the family of Lanczos polynomials determined by the spectral measure. As established in eq.~\eqref{eq:polynomial_orthogonality_TFD}, these polynomials satisfy
\begin{equation}
\frac{1}{D}
\sum_{i=0}^{D-1}
p_n(E_i)p_m(E_i)
=
\delta_{nm}.
\label{eq:polynomial_orthogonality_recalled}
\end{equation}
The dependence of the kernel on the spectrum is thus entirely encoded in the corresponding orthonormal-polynomial family. Unlike the spectral form factor, however, the kernel is not known in closed form for a generic random-matrix ensemble. Consequently, no closed-form expression for the spread complexity is presently available for such ensembles. One case in which the kernel can be constructed explicitly is that of a finite uniform lattice, and this observation will provide the starting point for our analysis.

The reference uniform lattice is defined by
\begin{equation}
\mathcal L_{\rm uni}
=
\{0,1,\ldots,D-1\}.
\label{eq:reference_uniform_lattice}
\end{equation}
The corresponding orthonormal polynomials are the discrete Chebyshev, or Gram, polynomials $t_n(x;D)$, which we normalize according to
\begin{equation}
\frac{1}{D}
\sum_{i=0}^{D-1}
t_n(i;D)t_m(i;D)
=
\delta_{nm},
\qquad
n,m=0,\ldots,D-1.
\label{eq:discrete_chebyshev_orthogonality}
\end{equation}
This normalization is identical to that of the Lanczos polynomials in eq.~\eqref{eq:polynomial_orthogonality_recalled} and, in particular, implies that the zeroth polynomial is
\begin{equation}
t_0(x;D)=1.
\end{equation}

We now formulate the central hypothesis of this work.
\begin{hypothesis}
\label{hyp:kernel_universality}
After unfolding, the normalized Krylov kernel associated with a generic spectrum, such as a GOE, GUE, GSE, or Poisson spectrum, is approximately described by the kernel constructed on a finite uniform lattice,
\begin{equation}
\mathcal K_{\rm uni}(i,j)
=
\frac{1}{D^2}
\sum_{n=0}^{D-1}
n\,t_n(i;D)t_n(j;D).
\label{eq:uniform_lattice_kernel}
\end{equation}
\end{hypothesis}

The discrete Chebyshev polynomials admit the hypergeometric representation
\begin{equation}
\begin{gathered}
t_n(x;D)
=
\mathcal N_n\,
{}_3F_2
\left(
\begin{matrix}
-n,\;n+1,\;-x\\
1,\;1-D
\end{matrix}
;1
\right),
\\
n=0,\ldots,D-1,
\end{gathered}
\label{eq:discrete_chebyshev_polynomial}
\end{equation}
where, with the normalization convention \eqref{eq:discrete_chebyshev_orthogonality}, the normalization constant is
\begin{equation}
\mathcal N_n
=
\sqrt{
(2n+1)
\frac{(D-1)!\,D!}
     {(D+n)!\,(D-n-1)!}
}.
\label{eq:discrete_chebyshev_normalization}
\end{equation}
For $n=0$, this gives
\begin{equation}
\mathcal N_0=1,
\end{equation}
and hence $t_0(x;D)=1$, in agreement with
eq.~\eqref{eq:discrete_chebyshev_orthogonality}.

The motivation for the hypothesis follows from the role of unfolding. Unfolding removes the smooth variation of the density of states and normalizes the local mean level spacing. An unfolded spectrum may therefore be compared naturally with a uniform lattice of unit mean spacing. Since the Krylov kernel is completely determined by the orthonormal-polynomial family associated with the spectral measure, this observation motivates comparing the kernels of unfolded spectra with the corresponding uniform-lattice kernel.

Let
\begin{equation}
\mathcal L_{\rm unfold}
=
\{\varepsilon_i\}_{i=0}^{D-1}
\label{eq:unfolded_lattice}
\end{equation}
denote an unfolded spectrum, ordered such that
\begin{equation}
\varepsilon_{i+1}>\varepsilon_i.
\end{equation}
We denote the corresponding Lanczos-polynomial family by $p_n^{(\rm unfold)}(x)$. With the normalization inherited from the infinite-temperature TFD state, these polynomials satisfy
\begin{equation}
\begin{gathered}
\frac{1}{D}
\sum_{i=0}^{D-1}
p_n^{(\rm unfold)}(\varepsilon_i)
p_m^{(\rm unfold)}(\varepsilon_i)
=
\delta_{nm},
\\
n,m=0,\ldots,D-1.
\end{gathered}
\label{eq:unfolded_orthogonality}
\end{equation}
The associated normalized Krylov kernel is therefore
\begin{equation}
\mathcal K_{\rm unfold}(i,j)
=
\frac{1}{D^2}
\sum_{n=0}^{D-1}
n\,
p_n^{(\rm unfold)}(\varepsilon_i)
p_n^{(\rm unfold)}(\varepsilon_j).
\label{eq:unfolded_kernel}
\end{equation}

After unfolding, the mean level spacing is unity. Up to an irrelevant overall translation, the unfolded levels may be parametrized as
\begin{equation}
\varepsilon_i
=
i+\delta_i,
\qquad
i=0,\ldots,D-1,
\label{eq:unfolded_near_lattice}
\end{equation}
where $\delta_i$ denotes the deviation of the $i$th unfolded level from the corresponding point of the reference uniform lattice. Equation \eqref{eq:unfolded_near_lattice} should be understood as the definition of $\delta_i$ and does not require these deviations to be small.

\begin{figure}
    \centering
    \includegraphics[width=0.7\linewidth]{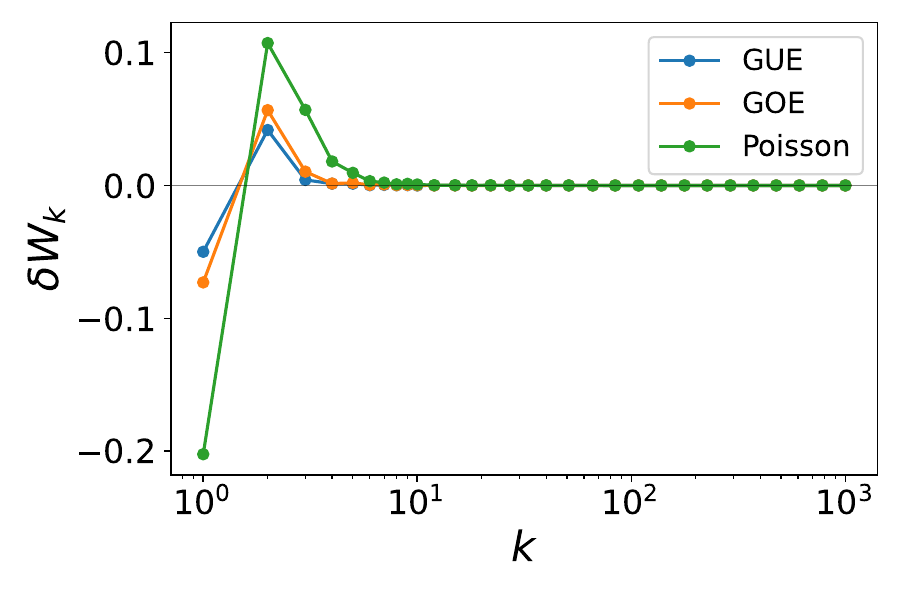}
    \caption{Normalized difference $\delta W_k$, Eq.~\eqref{eq:delta_W}, between the integrated diagonal weights of the unfolded and uniform-lattice Krylov kernels for single GUE, GOE and Poisson realizations at $D = 1000$. The discrepancy is confined to the first few diagonals and decays rapidly to zero with increasing $k$; its magnitude grows as level repulsion weakens, being largest for Poisson statistics.}
    \label{fig:deltaW_k}
\end{figure}

Why the kernels of the unfolded spectrum and of the uniform lattice should nearly agree, despite the deviations $\delta_i$, can be understood heuristically as follows. The Lanczos coefficients with $n \ll D$ are determined by the low moments of the spectral measure, $D^{-1}\sum_i \varepsilon_i^m$, which differ from those of the uniform lattice only by small relative corrections; correspondingly, polynomials of degree $n \ll D$ vary slowly on the scale of the level spacing and are insensitive to the deviations $\delta_i$. More generally, for discrete measures the leading asymptotics of the orthogonal polynomials, such as their zero distribution, are fixed for $n/D$ held constant by the limiting density of the nodes through a constrained equilibrium problem~\cite{Rakhmanov1996,DragnevSaff1997}, a discrete analogue of the Szeg\H{o}--Rakhmanov theory for continuous measures~\cite{Szego1939,Rakhmanov1977}. After unfolding, this density is uniform for any spectrum, as for the lattice, so that the positions of individual levels enter only at subleading order. In the bulk of the spectrum, they are resolved only by polynomials of degree comparable to $D$, which govern the kernel at short separations $|i-j|$; the residual discrepancy is therefore confined to the first few diagonals, consistent with Fig.~\ref{fig:deltaW_k}. Its size is set by the local level fluctuations: it grows as level repulsion weakens and is largest for Poisson spectra.

To compare the polynomial evaluations on the unfolded spectrum and on the uniform lattice, we introduce the matrices
\begin{equation}
\begin{gathered}
\left(O_{\rm unfold}\right)_{in}
=
\frac{1}{\sqrt D}\,
p_n^{(\rm unfold)}(\varepsilon_i),
\\
\left(O_{\rm uni}\right)_{in}
=
\frac{1}{\sqrt D}\,
t_n(i;D),
\\
i,n=0,\ldots,D-1.
\end{gathered}
\label{eq:polynomial_evaluation_matrices}
\end{equation}
The orthogonality relations
\eqref{eq:unfolded_orthogonality} and
\eqref{eq:discrete_chebyshev_orthogonality} imply
\begin{equation}
O_{\rm unfold}^{\mathsf T}O_{\rm unfold}
=
O_{\rm uni}^{\mathsf T}O_{\rm uni}
=
\mathbb I_D.
\label{eq:polynomial_matrix_orthogonality}
\end{equation}
Because the Krylov dimension is maximal, $K=D$, both matrices are square. They are therefore orthogonal and consequently also satisfy
\begin{equation}
O_{\rm unfold}O_{\rm unfold}^{\mathsf T}
=
O_{\rm uni}O_{\rm uni}^{\mathsf T}
=
\mathbb I_D.
\label{eq:polynomial_matrix_completeness}
\end{equation}

We define the overlap matrix between the two polynomial-evaluation matrices by
\begin{equation}
M
=
O_{\rm unfold}O_{\rm uni}^{\mathsf T}.
\label{eq:poly_overlap}
\end{equation}
Using eq.~\eqref{eq:polynomial_matrix_completeness}, one immediately obtains
\begin{equation}
O_{\rm unfold}
=
M O_{\rm uni}.
\label{eq:polynomial_basis_relation}
\end{equation}
Moreover,
\begin{align}
M M^{\mathsf T}
&=
O_{\rm unfold}
O_{\rm uni}^{\mathsf T}
O_{\rm uni}
O_{\rm unfold}^{\mathsf T}
=
\mathbb I_D,
\nonumber\\
M^{\mathsf T}M
&=
O_{\rm uni}
O_{\rm unfold}^{\mathsf T}
O_{\rm unfold}
O_{\rm uni}^{\mathsf T}
=
\mathbb I_D.
\label{eq:overlap_orthogonality}
\end{align}
Thus, $M$ is an orthogonal matrix acting in the ordered spectral-index space. It relates the polynomial evaluations on the uniform lattice to those on the unfolded spectrum.

After fixing the same sign convention for the leading coefficient of each polynomial, identical polynomial-evaluation matrices correspond to
\begin{equation}
M=\mathbb I_D.
\end{equation}
Deviations of $M$ from the identity therefore quantify the difference between the two polynomial-evaluation matrices. More generally, the localization properties of $M$ characterize the extent to which polynomial evaluations at different ordered spectral positions are mixed under the transformation from the uniform lattice to the unfolded spectrum.

For later convenience, we introduce the diagonal degree matrix
\begin{equation}
\Lambda
=
\operatorname{diag}(0,1,\ldots,D-1),
\label{eq:degree_matrix}
\end{equation}
whose diagonal entries are the polynomial degrees appearing in the definition of the Krylov kernel. Using eq.~\eqref{eq:polynomial_evaluation_matrices}, the normalized kernel matrices may be written as
\begin{align}
\mathcal K_{\rm unfold}
&=
\frac{1}{D}\,
O_{\rm unfold}
\Lambda
O_{\rm unfold}^{\mathsf T},
\label{eq:unfolded_kernel_matrix_form}
\\
\mathcal K_{\rm uni}
&=
\frac{1}{D}\,
O_{\rm uni}
\Lambda
O_{\rm uni}^{\mathsf T}.
\label{eq:uniform_kernel_matrix_form}
\end{align}
Substituting
eq.~\eqref{eq:polynomial_basis_relation}
into
eq.~\eqref{eq:unfolded_kernel_matrix_form},
we obtain
\begin{align}
\mathcal K_{\rm unfold}
&=
\frac{1}{D}
M
O_{\rm uni}
\Lambda
O_{\rm uni}^{\mathsf T}
M^{\mathsf T}
\nonumber\\
&=
M
\mathcal K_{\rm uni}
M^{\mathsf T}.
\end{align}
Hence,
\begin{equation}
\mathcal K_{\rm unfold}
=
M
\mathcal K_{\rm uni}
M^{\mathsf T}.
\label{eq:kernel_overlap_relation}
\end{equation}
Equation~\eqref{eq:kernel_overlap_relation} is exact. It shows that the difference between the Krylov kernel associated with the unfolded spectrum and that of the reference uniform lattice is entirely encoded in the overlap matrix $M$.

If the action of $M$ produces only small changes in the uniform-lattice kernel, and in particular if $M$ is sufficiently close to the identity, eq.~\eqref{eq:kernel_overlap_relation} immediately implies
\begin{equation}
\mathcal K_{\rm unfold}(i,j)
\simeq
\mathcal K_{\rm uni}(i,j).
\label{eq:kernel_universality_statement}
\end{equation}

\begin{figure}[tbp]
    \centering
    \includegraphics[width=0.7\linewidth]{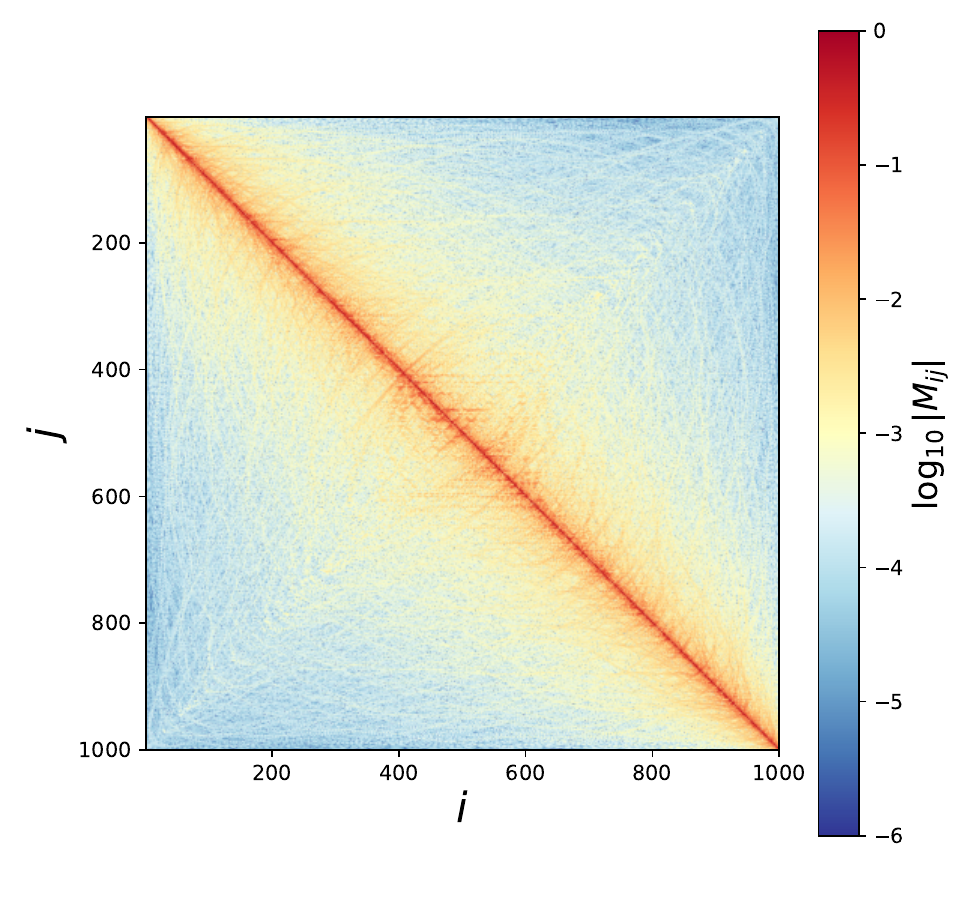}
    \caption{Heat map of $\log_{10}|M_{ij}|$ for the overlap matrix
$M=O_{\rm unfold}O_{\rm uni}^{\mathsf T}$ at $D=1000$, shown on a six-decade
diverging color scale. The matrix is sharply concentrated along the
principal diagonal, where $|M_{ij}|\sim 1$, and decays by several
orders of magnitude within a narrow band around it, indicating that the
transformation between the uniform-lattice and unfolded
polynomial-evaluation matrices is approximately local in the ordered
spectral index.}
    \label{fig:M_ij_matrixplot}
\end{figure}

The remainder of this section is devoted to testing this hypothesis numerically. We first examine the overlap matrix $M$ itself. Fig.~\ref{fig:M_ij_matrixplot} shows a heat map of $M$ for a representative realization. The matrix elements are strongly concentrated along the principal diagonal $i=j$, while the off-diagonal entries decay rapidly away from it.
\begin{figure}[tbp]
    \centering
    \includegraphics[width=1\linewidth]{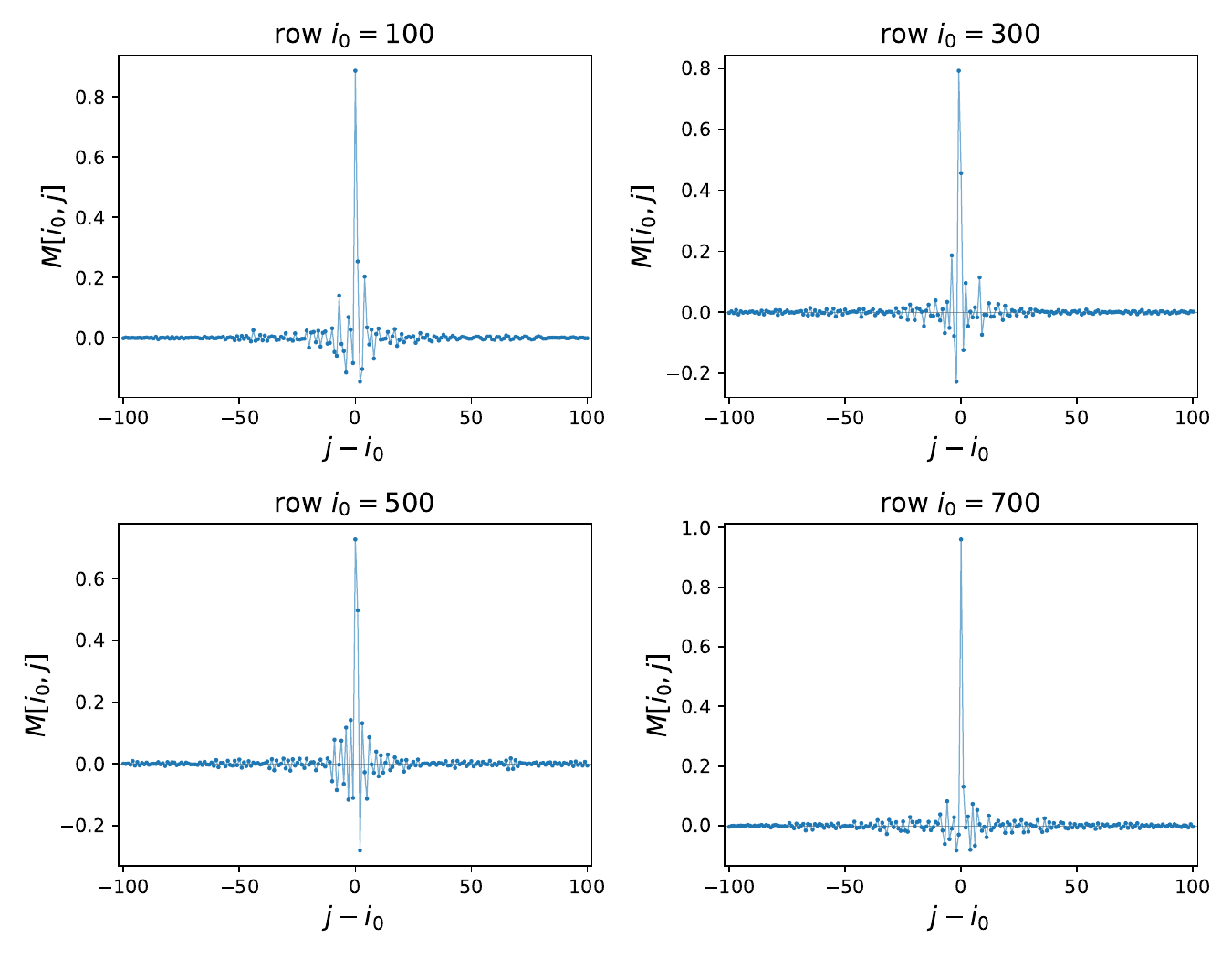}
    \caption{Row profiles $M[i_0,j]$ plotted against $j-i_0$ for $i_0 = 100, 300, 500, 700$ ($D = 1000$); markers denote integer $j$. Each row is sharply peaked at $j = i_0$, with rapidly decaying oscillatory tails, showing that the localization of $M$ persists throughout the bulk of the spectrum.}
    \label{fig:M_different_diagonal}
\end{figure}
\begin{figure}[tbp]
    \centering
    \includegraphics[width=1\linewidth]{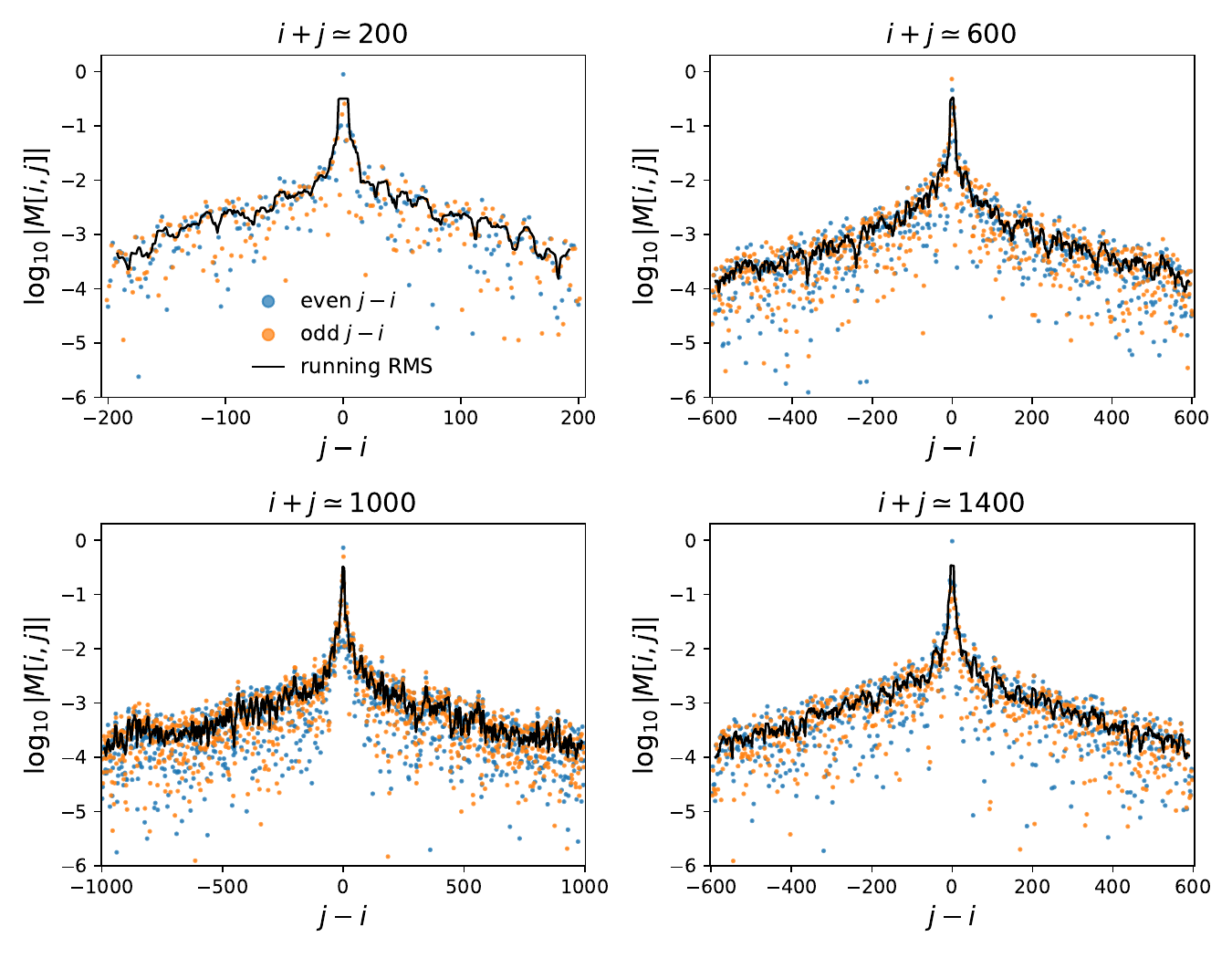}
    \caption{Anti-diagonal slices $\log_{10}|M[i,j]|$ plotted against $j-i$ along $i+j \simeq s$, with $s = 200, 600, 1000, 1400$, crossing the principal diagonal at the rows of Fig.~\ref{fig:M_different_diagonal} ($D = 1000$). Even (blue) and odd (orange) offsets are taken from $i+j=s$ and $s+1$, respectively; the black line is a running RMS over nine offsets. The local bumps in the tail correspond to the ridges in Fig.~\ref{fig:M_ij_matrixplot}.}
    \label{fig:M_anti_diagonal}
\end{figure}

The localization of $M$ is further illustrated in Fig.~\ref{fig:M_different_diagonal}, which displays representative row profiles of $M$. In every case, the dominant weight is concentrated near $j=i_0$, while the matrix elements decay rapidly with increasing $|j-i_0|$, indicating that the transformation between the two polynomial-evaluation matrices is approximately local in the ordered spectral index.

To resolve the structure of $M$ transverse to the principal diagonal, Fig.~\ref{fig:M_anti_diagonal} shows $\log_{10}|M[i,j]|$ along anti-diagonals $i+j \simeq \mathrm{const}$, which probe the dependence on the separation $j-i$ at a fixed position in the spectrum. Each slice consists of a sharp central peak followed by a weak tail that decays slowly with $|j-i|$; the faint ridges visible in Fig.~\ref{fig:M_ij_matrixplot} appear as local bumps on this tail. The tail nevertheless carries little weight and the transformation between the two polynomial-evaluation matrices is therefore local up to a small remainder.

\begin{figure}
    \centering
    \includegraphics[width=1\linewidth]{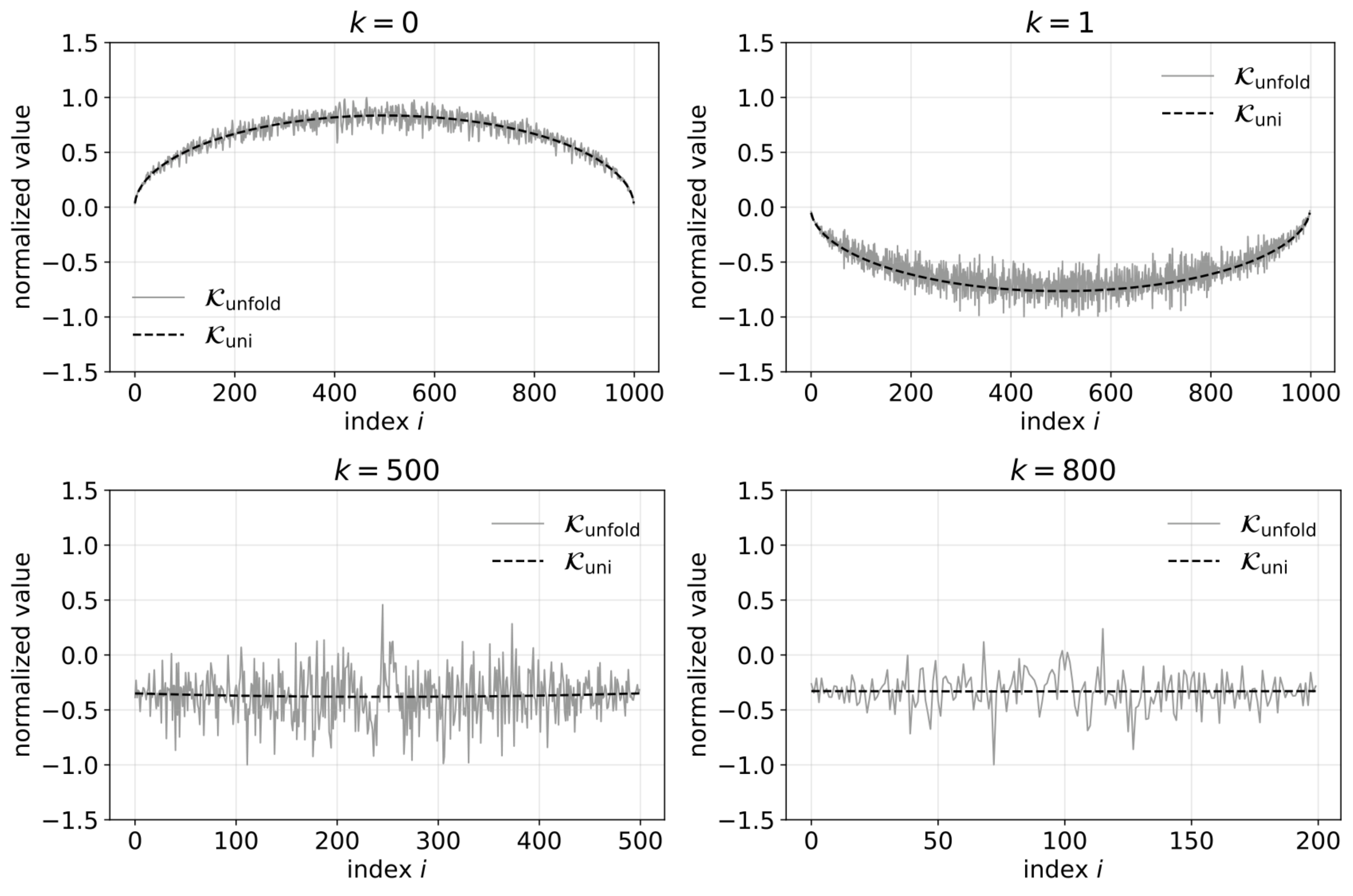}
    \caption{Diagonal profile
    $\mathcal K(i,i+k)$
    for
    $k=0,1,500,800$
    at
    $D=1000$,
    comparing a single unfolded GUE realization (grey) with the analytic
    uniform-lattice kernel.
    The curves are normalized by
    $\max_i |\mathcal K_{\rm unfold}(i,i+k)|$.}
    \label{fig:comparison_Kernel}
\end{figure}

As a complementary test of the universality hypothesis, Fig.~\ref{fig:comparison_Kernel} compares the diagonal profiles $\mathcal K_{\rm unfold}(i,i+k)$ obtained from a single unfolded GUE realization with the analytic uniform-lattice prediction $\mathcal K_{\rm uni}(i,i+k)$ for several representative values of $k$. Despite the fluctuations present in the unfolded kernel, the analytic uniform-lattice kernel accurately follows its smooth envelope over the entire range shown.

As a further quantitative comparison, we define the normalized difference
\begin{equation}
\delta W_k
=
\frac{(W_k)_{\rm unfold}
-
(W_k)_{\rm uni}}
{(W_1)_{\rm uni}},
\label{eq:delta_W}
\end{equation}
where $W_k$ is the integrated diagonal weight introduced in eq.~\eqref{eq:W_k_defn}. Fig.~\ref{fig:deltaW_k} shows that $\delta W_k$ is appreciable only for the first few diagonals and rapidly decays to values that are essentially indistinguishable from zero for all three ensembles. Thus, although the unfolded and uniform kernels are not identical, their integrated diagonal weights differ only on the first few diagonals, and only slightly for the Gaussian ensembles. For the GUE and GOE, we have verified that $\delta W_k$ at fixed small $k$ does not decrease with $D$: the uniform-lattice kernel is therefore not asymptotically exact, but retains a small, nearly $D$-independent discrepancy confined to the first few diagonals, reflecting the $O(1)$ spacing fluctuations of the unfolded spectrum.

The same checks for Poisson spectra, which lack spectral rigidity, are presented in Appendix~\ref{app:poisson_checks}. The overlap matrix $M$ remains concentrated in a band around the principal diagonal, but this band is broader than for the GUE, and the row maxima are displaced from $j = i_0$ by a few sites, reflecting the larger deviations $\delta_i$ of the unfolded Poisson levels. Correspondingly, the discrepancy $\delta W_k$ is larger and extends over more diagonals (Fig.~\ref{fig:deltaW_k}). Nevertheless, the diagonal expansion~\eqref{eq:complexity_sum1} converges rapidly, and the uniform-lattice kernel reproduces the qualitative behavior of the Poisson spread complexity, with small quantitative deviations in the growth regime
(Sec.~\ref{sec:Poisson_no_peak}).

Taken together, the localization of the overlap matrix $M$, the close agreement of the diagonal kernel profiles, and the rapid decay of $\delta W_k$ provide strong numerical support for the approximate relation
\begin{equation}
\mathcal K_{\rm unfold}(i,j)
\simeq
\mathcal K_{\rm uni}(i,j),
\end{equation}
which forms the basis for the analytic treatment of the spread complexity presented in the following sections.
%

\section{Spectral origin of the spread complexity peak}
\label{sec:analytic_peak}

Having established the approximate kernel universality in the previous section, we now derive an analytic approximation to the ensemble-averaged spread complexity for the Gaussian random-matrix ensembles. We begin with the exact kernel representation,
\begin{equation}
C(t)
=
\sum_{i,j=0}^{D-1}
\mathcal K(i,j)
e^{-i(E_i-E_j)t},
\label{eq:C_exact_again}
\end{equation}
which, after unfolding the spectrum, can be reorganized according to the level-index separation $k=j-i$. This gives
\begin{equation}
\begin{gathered}
C_{\rm unfold}(t)
=
\sum_{i=0}^{D-1}
\mathcal K_{\rm unfold}(i,i)
\\
+
2
\sum_{k=1}^{D-1}
\sum_{i=0}^{D-1-k}
\mathcal K_{\rm unfold}(i,i+k)
\cos\!\left[
(\varepsilon_{i+k}-\varepsilon_i)t
\right].
\end{gathered}
\label{eq:C_unfold_exact}
\end{equation}
Upon ensemble averaging, the oscillatory factor is replaced by its average over the $k$th-nearest-neighbour spacing distribution,
\begin{equation}
\left\langle
\cos\!\left[
(\varepsilon_{i+k}-\varepsilon_i)t
\right]
\right\rangle
=
\int_0^\infty ds\,p_k(s)\cos(st)
\equiv
\Phi_k(t),
\label{eq:Phi_def}
\end{equation}
where $p_k(s)$ denotes the probability density of the unfolded $k$th-nearest-neighbour spacing. Substituting the approximate kernel
$\mathcal K_{\rm unfold}\simeq\mathcal K_{\rm uni}$ into eq.~\eqref{eq:C_unfold_exact} and using eq.~\eqref{eq:Phi_def}, we obtain
\begin{equation}
\begin{gathered}
\left\langle
C_{\rm unfold}(t)
\right\rangle
\simeq
\sum_{i=0}^{D-1}
\mathcal K_{\rm uni}(i,i)
\\
+
2
\sum_{k=1}^{D-1}
\Phi_k(t)
\sum_{i=0}^{D-1-k}
\mathcal K_{\rm uni}(i,i+k).
\end{gathered}
\label{eq:C_factorized}
\end{equation}

Substituting the large-$D$ expressions for the diagonal sums of the uniform kernel (see appendix \ref{app:uniform_kernel_sums}),
\begin{equation}
\begin{gathered}
\sum_{i=0}^{D-1}
\mathcal K_{\rm uni}(i,i)
\sim
\frac{D}{2},
\\ 
\sum_{i=0}^{D-1-k}
\mathcal K_{\rm uni}(i,i+k)
\sim
-
\frac{D}
{2(4k^2-1)},
\end{gathered}
\label{eq:Kernel_sum}
\end{equation}
we obtain, in the large-$D$ limit,
\begin{equation}
\frac{\langle C_{\rm unfold}(t) \rangle}{D}
\simeq
\frac12 \left(1
-
2\sum_{k=1}^{\infty}
\frac{\Phi_k(t)}
{4k^2-1} \right).
\label{eq:master_formula}
\end{equation}

Equation~\eqref{eq:master_formula} provides the central analytic result of this section. It expresses the spread complexity entirely in terms of the universal kernel coefficients and the characteristic functions of the unfolded level-spacing distributions. Consequently, all ensemble dependence enters solely through the spacing statistics of the unfolded spectrum.

It is important to emphasize that the variable $t$ in Eq.~\eqref{eq:master_formula} is Fourier-conjugate to the unfolded energies $\{\varepsilon_i\}$, rather than to the physical energies $\{E_i\}$. To avoid confusion, we denote this variable by $\tau$ for the remainder of this section. With this convention, the relevant complexity is that of the unfolded Hamiltonian $H_{\rm unfold} = N(H)$, where $N(E)$ is the smooth counting function (the mean number of levels below $E$) used for unfolding, so that $\varepsilon_i \simeq N(E_i)$; its eigenvalues are precisely the unfolded energies. We explore the relation between folded and unfolded complexities in Appendix~\ref{app:folded_unfolded_time}.

In the following, we investigate the emergence of the spread complexity peak for the Gaussian random-matrix ensembles. To obtain simple analytic expressions, we retain only the leading contribution in the diagonal expansion, namely the nearest-neighbour term ($k=1$). This approximation is motivated by the rapid suppression of the higher-order contributions. Indeed, the contribution from the $k$th-neighbour spacing is weighted by the kernel coefficient $(4k^2-1)^{-1}$. More precisely, after retaining the first $m$ diagonal contributions, the remaining terms satisfy the uniform bound
\begin{equation}
\left|
\sum_{k=m+1}^{\infty}
\frac{\Phi_k(\tau)}{4k^2-1}
\right|
\leq
\sum_{k=m+1}^{\infty}
\frac{1}{4k^2-1}
=
\frac{1}{2(2m+1)},
\label{eq:tail_bound}
\end{equation}
where we have used the inequality $|\Phi_k(\tau)|\leq1$. This establishes the controlled convergence of the diagonal expansion and provides a quantitative justification for using the leading ($k=1$) contribution as the first analytic approximation to the complexity peak.

With this, the leading time-dependent contribution is therefore
\begin{equation}
\frac{\langle C_{\rm unfold}^{(1)}(\tau) \rangle}{D}
=
\frac{1}{2}
\left(
1-\frac{2}{3}\Phi_1(\tau)
\right).
\label{eq:leading_spacing_approximation}
\end{equation}
This approximation is designed to capture the peak. It does not reproduce the exact initial condition, since $\Phi_1(0)=1$ implies
\begin{equation}
\frac{\langle C_{\rm unfold}^{(1)}(0) \rangle}{D}
=
\frac{1}{6},
\end{equation}
whereas the exact complexity satisfies $C_{\rm unfold}(0)=0$. This condition is recovered only after summing over all values of $k$. This is consistent with Fig.~\ref{fig:Krylov_comparison}, where the truncated approximations $C^{(m)}(t)$ are shown to have non-zero initial values at $t=0$, which decrease as additional terms are included. 

For notational convenience, we suppress the ensemble-average brackets $\langle\cdots\rangle$ in the following. Thus, unless stated otherwise, $C(\tau)$ denotes the ensemble-averaged spread complexity.

\subsection*{Gaussian Unitary Ensemble (GUE)}

For the GUE, the nearest-neighbour spacing distribution is approximated by the Wigner surmise
\begin{equation}
p_1^{\rm GUE}(s)
=
\frac{32}{\pi^2}
s^2
e^{-4s^2/\pi},
\qquad
s\geq0.
\label{eq:GUE_wigner_surmise}
\end{equation}
This distribution is normalized and has unit mean spacing:
\begin{equation}
\int_0^\infty ds\,p_1^{\rm GUE}(s)=1,
\qquad
\int_0^\infty ds\,s\,p_1^{\rm GUE}(s)=1.
\end{equation}
Its cosine transform is
\begin{align}
\Phi_1^{\rm GUE}(\tau)
&=
\frac{32}{\pi^2}
\int_0^\infty ds\,
s^2e^{-4s^2/\pi}\cos(s\tau) \nonumber \\
&= \left(
1-\frac{\pi\tau^2}{8}
\right)
e^{-\pi\tau^2/16}.
\label{eq:GUE_cosine_transform_start}
\end{align}
Substitution into eq.~\eqref{eq:leading_spacing_approximation} gives
\begin{equation}
\frac{C_{\rm GUE}^{(1)}(\tau)}{D}
=
\frac{1}{2}
-
\frac{1}{3}
\left(
1-\frac{\pi\tau^2}{8}
\right)
e^{-\pi\tau^2/16}.
\label{eq:GUE_Krylov_analytic}
\end{equation}

The maximum of the complexity corresponds to the minimum of $\Phi_1^{\rm GUE}(\tau)$ which is located at
\begin{equation}
\tau_{\rm peak}^{\rm GUE}
=
\sqrt{\frac{24}{\pi}} \approx 2.764 \, .
\label{eq:GUE_peak_time}
\end{equation}
At this point we have
\begin{equation}
\frac{C_{\rm GUE,peak}^{(1)}}{D}
=
\frac{1}{2}
+
\frac{2}{3}e^{-3/2} \approx 0.649.
\label{eq:GUE_peak_height}
\end{equation}

\subsection*{Gaussian Orthogonal Ensemble (GOE)}

For the GOE we have
\begin{equation}
p_1^{\rm GOE}(s)
=
\frac{\pi}{2}
s\,e^{-\pi s^2/4},
\qquad
s\geq0.
\label{eq:GOE_wigner_surmise}
\end{equation}
Its cosine transform is
\begin{align}
\Phi_1^{\rm GOE}(\tau)
&=
\frac{\pi}{2}
\int_0^\infty ds\,
s\,e^{-\pi s^2/4}\cos(s\tau) \nonumber \\
& = 1
-
\frac{2\tau}{\sqrt\pi}
F\left(
\frac{\tau}{\sqrt\pi}
\right).
\label{eq:GOE_cosine_transform_start}
\end{align}
where the Dawson
function $F(x)$ is defined as
\begin{equation}
F(x)
=
e^{-x^2}
\int_0^x du\,e^{u^2}.
\label{eq:Dawson_definition}
\end{equation}
The leading approximation to the GOE spread complexity is therefore
\begin{equation}
\frac{C_{\rm GOE}^{(1)}(\tau)}{D}
=
\frac{1}{2}
-
\frac{1}{3}
\left[
1
-
\frac{2\tau}{\sqrt\pi}
F\left(
\frac{\tau}{\sqrt\pi}
\right)
\right].
\label{eq:GOE_Krylov_analytic}
\end{equation}

To determine the peak, introduce
\begin{equation}
x=\frac{\tau}{\sqrt\pi}.
\end{equation}
Equation~\eqref{eq:GOE_cosine_transform_start} then reads
\begin{equation}
\Phi_1^{\rm GOE}(x)
=
1-2xF(x).
\end{equation}
The Dawson function satisfies
\begin{equation}
F'(x)=1-2xF(x).
\label{eq:Dawson_derivative}
\end{equation}
Therefore,
\begin{align}
\frac{d\Phi_1^{\rm GOE}}{dx}
&=
-2F(x)-2xF'(x)
\nonumber\\
&=
-2F(x)-2x+4x^2F(x).
\end{align}
The nontrivial extremum is determined by
\begin{equation}
F(x_{\rm peak})
=
\frac{x_{\rm peak}}{2x_{\rm peak}^2-1}.
\label{eq:GOE_peak_equation}
\end{equation}
This equation is transcendental and must be solved numerically. Its relevant positive solution is
\begin{equation}
x_{\rm peak}\simeq1.502,
\end{equation}
which gives
\begin{equation}
\tau_{\rm peak}^{\rm GOE}
=
\sqrt\pi\,x_{\rm peak}
\simeq2.662.
\label{eq:GOE_peak_time}
\end{equation}
The corresponding peak height is
\begin{equation}
\frac{C_{\rm GOE,peak}^{(1)}}{D}
=
\frac{1}{2}
-
\frac{1}{3}
\left[
1-2x_{\rm peak}F(x_{\rm peak})
\right]
\simeq0.595.
\label{eq:GOE_peak_height}
\end{equation}

Equations~\eqref{eq:GUE_Krylov_analytic} and \eqref{eq:GOE_Krylov_analytic} show explicitly how the random-matrix symmetry class enters the Krylov dynamics through the nearest-neighbour spacing distribution. In particular, the stronger level repulsion of the
GUE produces a more negative minimum of $\Phi_1(\tau)$ and, consequently, a higher spread complexity peak than in the GOE.

We stress once again that $\tau_{\rm peak}^{\rm GUE}$ and $\tau_{\rm peak}^{\rm GOE}$ are peak positions in unfolded time. They are not, in general, equal to the peak positions measured using the original Hamiltonian and the physical time variable $t$ (see appendix \ref{app:folded_unfolded_time}).

\begin{figure}
    \centering
    \includegraphics[width=1.0\linewidth]{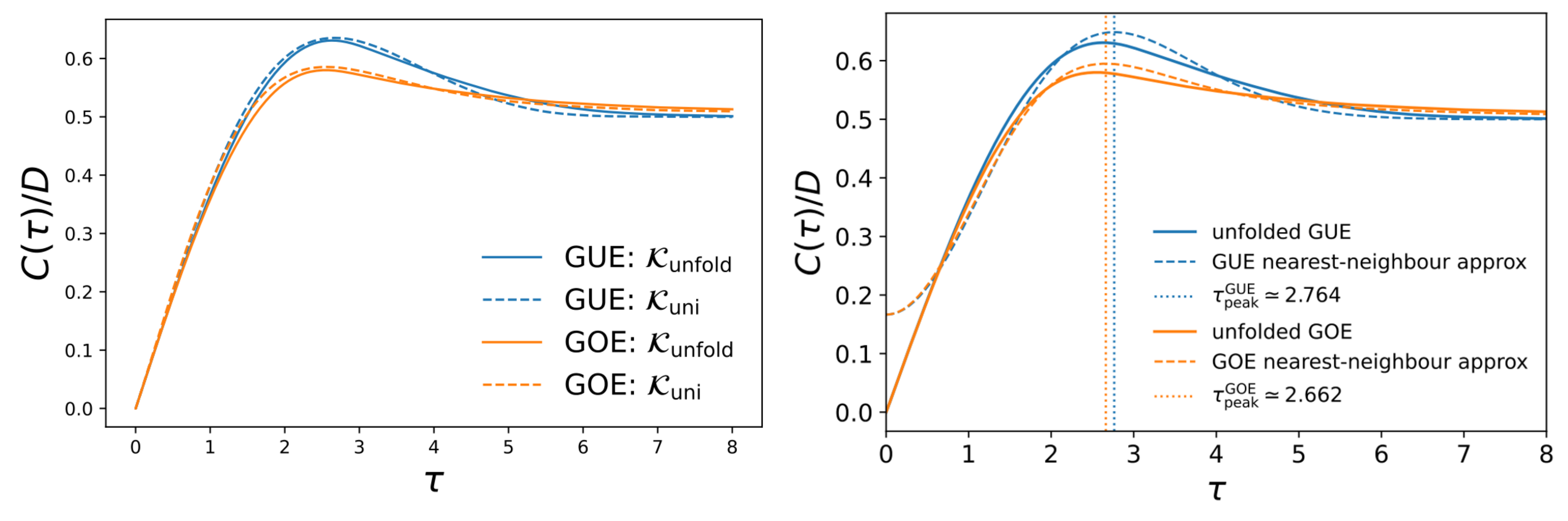}
    \caption{Left: ensemble-averaged $C(\tau)/D$ for GUE (blue) and GOE
    (orange), comparing the exact unfolded kernel
    $\mathcal{K}_{\rm unfold}$ (solid) with the uniform-lattice
    approximation $\mathcal{K}_{\rm uni}$ evaluated at the same unfolded
    energies (dashed), each averaged over 20 realizations. Right: the
    same ensemble-averaged $C(\tau)/D$ (solid) compared with the
    leading-order nearest-neighbour analytic approximation (dashed),
    with vertical dotted lines marking the corresponding analytic peak
    positions, $\tau_{\rm peak}^{\rm GUE}\simeq2.764$ and
    $\tau_{\rm peak}^{\rm GOE}\simeq2.662$.}
    \label{fig:complexity_GOE_GUE}
\end{figure}

To assess the accuracy of the universal-kernel approximation at the level of the complexity, we compare in Fig.~\ref{fig:complexity_GOE_GUE} the complexity computed using the exact unfolded kernel $\mathcal{K}_{\rm unfold}$ with that obtained from the uniform lattice kernel $\mathcal{K}_{\rm uni}$. The agreement is excellent for both the GOE and GUE ensembles, with the two curves being nearly indistinguishable over the entire time evolution, including the growth, peak, and plateau regimes. This provides strong numerical evidence that the universal-kernel hypothesis accurately captures the spread complexity itself.

As a further test of the leading-order approximation, the right panel of Fig.~\ref{fig:complexity_GOE_GUE} compares the full ensemble-averaged unfolded complexity with the nearest-neighbour analytic approximations of eqs.~\eqref{eq:GUE_Krylov_analytic} and
\eqref{eq:GOE_Krylov_analytic}, for GUE and GOE respectively. The vertical dotted lines mark the corresponding analytic peak positions, $\tau_{\rm peak}^{\rm GUE}\simeq2.764$ and $\tau_{\rm peak}^{\rm GOE}\simeq2.662$. The numerically observed peaks for both ensembles occur close to these predicted locations, confirming that retaining only the leading ($k=1$) diagonal already captures the position and approximate height of the complexity peak. 
%

\section{Absence of the complexity peak for Poisson statistics}
\label{sec:Poisson_no_peak}

For an unfolded Poisson spectrum, the spacings between consecutive levels are independent exponentially distributed random variables. Consequently, the spacing between two levels separated by $k$ indices follows the Gamma distribution
\begin{equation}
p_k^{\rm P}(s)
=
\frac{s^{k-1}e^{-s}}{(k-1)!},
\qquad
s\geq0.
\label{eq:Poisson_k_spacing}
\end{equation}
The corresponding cosine transform is
\begin{align}
\Phi_k^{\rm P}(\tau)
&=
\int_0^\infty ds\,
p_k^{\rm P}(s)\cos(s\tau)
\nonumber\\
&=
\operatorname{Re}
\left[
\frac{1}{(k-1)!}
\int_0^\infty ds\,
s^{k-1}e^{-(1-i\tau)s}
\right]
\nonumber\\
&=
\operatorname{Re}
\left(1-i\tau\right)^{-k}.
\label{eq:Poisson_k_characteristic}
\end{align}
Equivalently,
\begin{equation}
\Phi_k^{\rm P}(\tau)
=
\frac{
\cos\!\left(k\arctan\tau\right)
}{
(1+\tau^2)^{k/2}
}.
\label{eq:Poisson_k_characteristic_real}
\end{equation}
Although the individual functions $\Phi_k^{\rm P}(\tau)$ are not necessarily positive for $k\geq2$, the complete sum over all neighbour separations can be performed exactly. In the large-$D$ limit, the complexity is
\begin{equation}
\frac{C_{\rm P}(\tau)}{D}
=
\frac12
-
\sum_{k=1}^{\infty}
\frac{\Phi_k^{\rm P}(\tau)}{4k^2-1}.
\label{eq:Poisson_full_complexity}
\end{equation}
Introducing
\begin{equation}
z(\tau)
=
\frac{1}{1-i\tau},
\end{equation}
such that
\begin{equation}
\Phi_k^{\rm P}(\tau)
=
\operatorname{Re}\!\left[z(\tau)^k\right],
\end{equation}
and using the integral representation
\begin{equation}
\frac{1}{4k^2-1}
=
\frac12
\int_0^1 dx\,
(1-x^2)x^{2k-2},
\label{eq:kernel_coefficient_integral}
\end{equation}
the infinite series appearing in
eq.~\eqref{eq:Poisson_full_complexity} becomes
\begin{align}
\sum_{k=1}^{\infty}
\frac{\Phi_k^{\rm P}(\tau)}{4k^2-1}
&=
\frac12
\operatorname{Re}
\int_0^1 dx\,
(1-x^2)
\sum_{k=1}^{\infty}
z(\tau)^k x^{2k-2}
\nonumber\\
&=
\frac12
\operatorname{Re}
\int_0^1 dx\,
(1-x^2)
\frac{z(\tau)}{1-z(\tau)x^2}.
\label{eq:Poisson_sum_intermediate}
\end{align}
Since
\begin{equation}
\frac{z(\tau)}{1-z(\tau)x^2}
=
\frac{1}{1-x^2-i\tau},
\end{equation}
its real part is
\begin{equation}
\operatorname{Re}
\left[
\frac{1}{1-x^2-i\tau}
\right]
=
\frac{1-x^2}
{(1-x^2)^2+\tau^2}.
\end{equation}
It follows that
\begin{equation}
\sum_{k=1}^{\infty}
\frac{\Phi_k^{\rm P}(\tau)}{4k^2-1}
=
\frac12
\int_0^1 dx\,
\frac{(1-x^2)^2}
{(1-x^2)^2+\tau^2}.
\label{eq:Poisson_resummed_series}
\end{equation}
Substituting this result into eq.~\eqref{eq:Poisson_full_complexity} gives the exact resummed expression
\begin{equation}
\frac{C_{\rm P}(\tau)}{D}
=
\frac12
\int_0^1 dx\,
\frac{\tau^2}
{(1-x^2)^2+\tau^2}.
\label{eq:Poisson_resummed_complexity}
\end{equation}
This representation makes the absence of a complexity peak manifest. For every $x\in[0,1]$ and finite $\tau\geq0$,
\begin{equation}
0
\leq
\frac{\tau^2}
{(1-x^2)^2+\tau^2}
\leq
1,
\end{equation}
and therefore
\begin{equation}
0
\leq
\frac{C_{\rm P}(\tau)}{D}
<
\frac12
\qquad
\text{for finite }\tau.
\label{eq:Poisson_complexity_bound}
\end{equation}
The upper bound is approached only asymptotically:
\begin{equation}
\frac{C_{\rm P}(0)}{D}=0,
\qquad
\lim_{\tau\rightarrow\infty}
\frac{C_{\rm P}(\tau)}{D}
=
\frac12.
\label{eq:Poisson_limits}
\end{equation}
Moreover, differentiating eq.~\eqref{eq:Poisson_resummed_complexity} yields
\begin{equation}
\frac{d}{d\tau}
\left[
\frac{C_{\rm P}(\tau)}{D}
\right]
=
\int_0^1 dx\,
\frac{
\tau(1-x^2)^2
}{
\left[(1-x^2)^2+\tau^2\right]^2
}.
\label{eq:Poisson_derivative}
\end{equation}
For $\tau>0$, the integrand is non-negative and is strictly positive except at the endpoint $x=1$. Hence,
\begin{equation}
\frac{dC_{\rm P}(\tau)}{d\tau}>0,
\qquad
\tau>0.
\label{eq:Poisson_monotonicity}
\end{equation}
The Poisson complexity therefore increases monotonically from zero and approaches the plateau $D/2$ from below, without developing a finite-time maximum. This behaviour contrasts with the leading-order analysis of the Gaussian ensembles presented above, where the nearest-neighbour contribution already gives rise to a pronounced complexity peak.

\begin{figure}
\centering
\includegraphics[width=0.7\linewidth]{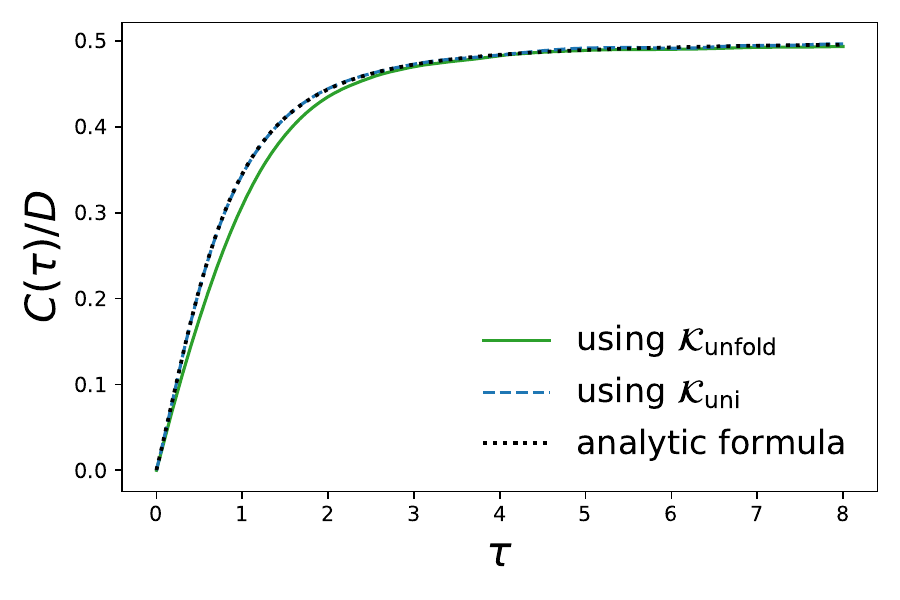}
\caption{Ensemble-averaged unfolded Poisson complexity $C(\tau)/D$
(green), compared with the uniform-kernel approximation evaluated at
the same energies (blue, dashed) and the exact resummed analytic
formula of eq.~\eqref{eq:Poisson_resummed_complexity} (black, dotted). The
averaging is done over 20 realizations at $D=1000$.}
\label{fig:folded_unfolded_poisson}
\end{figure}

Fig.~\ref{fig:folded_unfolded_poisson} confirms this numerically: the ensemble-averaged unfolded Poisson complexity, the uniform-kernel approximation, and the exact resummed formula of eq.~\eqref{eq:Poisson_resummed_complexity} all increase monotonically toward the plateau without a peak, and agree well, with small deviations in the growth regime.

\section{An Example: SYK Model}
\label{sec:syk}

The analysis so far has been carried out for Gaussian random-matrix ensembles and for Poisson statistics. Both are spectral models: they specify the statistics of the eigenvalues directly, without reference to an underlying Hamiltonian. It is therefore natural to ask whether the structures identified above---the banded Krylov kernel, the $k^{-2}$ hierarchy of integrated diagonal weights, and the approximate kernel universality of section~\ref{sec:the_hypothesis}---persist for a genuine interacting many-body system, whose Hamiltonian is fixed by a physical prescription and whose couplings are far fewer in number than the $O(D^2)$ independent entries of a random matrix. In this section we test this for the Sachdev--Ye--Kitaev model~\cite{Sachdev_1993, RevModPhys.94.035004}.

We consider the Majorana SYK model with $q=4$,
\begin{equation}
  H \;=\; \sum_{i<j<k<l} J_{ijkl}\,\chi_i\chi_j\chi_k\chi_l ,
  \qquad \{\chi_a,\chi_b\}=\delta_{ab},
  \label{eq:syk_H}
\end{equation}
where the $N$ Majorana fermions act on a Hilbert space of dimension $2^{N/2}$ and the couplings $J_{ijkl}$ are independent real Gaussian variables with
\begin{equation}
  \langle J_{ijkl}\rangle = 0, \qquad
  \langle J_{ijkl}^{2}\rangle = \frac{3!\,J^{2}}{N^{3}} .
  \label{eq:syk_variance}
\end{equation}
Since $q$ is even, $H$ commutes with the fermion parity operator, and the Hilbert space splits into two sectors of dimension $2^{N/2-1}$. The Krylov construction of section~\ref{sec:the_hypothesis} requires a nondegenerate spectrum, so we work throughout within a single parity sector; we take the positive-parity (even fermion number) sector. The symmetry class of each sector depends on $N \bmod 8$~\cite{Cotler:2016fpe, Garcia-Garcia:2016mno}: it is GOE for $N\equiv 0$, GUE for $N\equiv 2,6$, and GSE for $N\equiv 4$, the last case carrying a twofold Kramers degeneracy of every level. We choose
\begin{equation}
  N = 22 , \qquad D = 2^{N/2-1} = 1024 ,
\end{equation}
which lies in the GUE class and gives a sector dimension close to the $D=1000$ used for the random-matrix ensembles above, allowing a direct comparison\footnote{Spread complexity in the SYK model has been examined previously in~\cite{Balasubramanian:2022tpr, Erdmenger:2023wjg, Baggioli:2024wbz, Huh:2024ytz}; our interest here is in the structure of the underlying Krylov kernel rather than in the complexity profile itself.}.

\subsection*{Banded structure of the Kernel}

\begin{figure}
    \centering
    \includegraphics[width=1\linewidth]{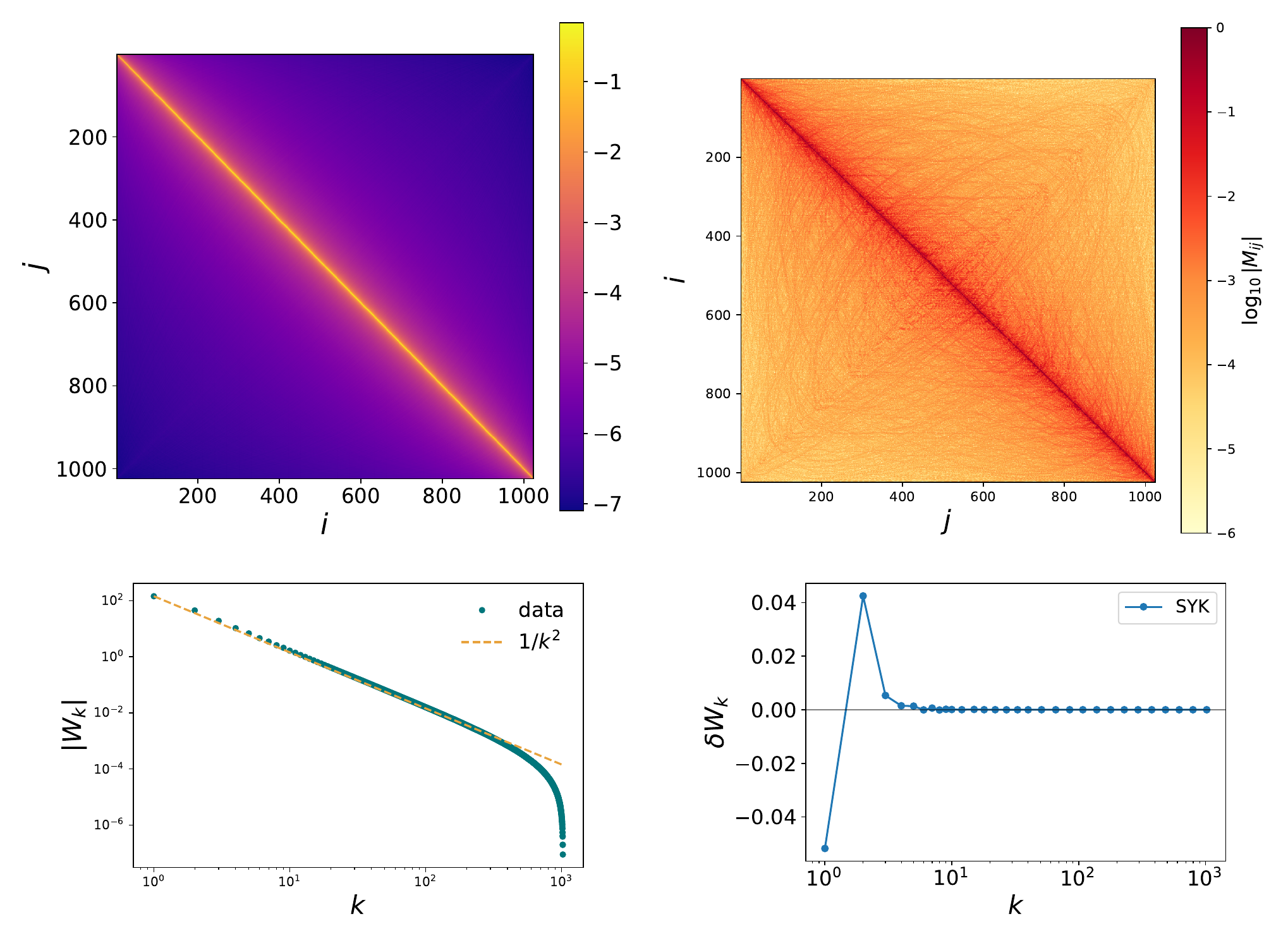}
    \caption{Krylov kernel and overlap matrix for the Majorana SYK model with $N=22$ fermions in the positive-parity sector, $D=1024$. Top left: $\log_{10}(\langle|\mathcal{K}(i,j)|\rangle)$ for the raw spectrum ($50$ realizations), as in Fig.~\ref{fig:K_ij_matrixplot}. Top right: $\log_{10}|M_{ij}|$ for a single unfolded realization, as in Fig.~\ref{fig:M_ij_matrixplot}. Bottom left: the integrated weight $|W_k|$ and its $1/k^2$ decay, as in Fig.~\ref{fig:Wk_vs_k}. Bottom right: the normalized difference $\delta W_k$ ($10$ realizations), as in Fig.~\ref{fig:deltaW_k}. The SYK model reproduces all four features found for the Gaussian ensembles.}
    \label{fig:four_figures_SYK}
\end{figure}

The top-left panel of Fig.~\ref{fig:four_figures_SYK} shows the heat map of $\log_{10}\langle|\mathcal{K}(i,j)|\rangle$ for the raw SYK spectrum. The kernel is concentrated in a narrow band around the principal diagonal and decays  toward the corners, reproducing the structure found for GUE in Fig.~\ref{fig:K_ij_matrixplot}. 

The integrated off-diagonal weights $W_k$ of eq.~\eqref{eq:W_k_defn} are negative for every $k$, and the bottom-left panel of Fig.~\ref{fig:four_figures_SYK} shows that $|W_k|$ follows the same $k^{-2}$ decay found for the Gaussian ensembles. Fitting $|W_k|=A\,k^{-\alpha}$ over $k\in[5,100]$, with uncertainties obtained by bootstrapping over disorder realizations, gives
\begin{equation}
  \alpha_{\rm SYK} = 2.0175 \pm 0.0005 ,
\end{equation}
to be compared with $\alpha_{\rm GUE}=2.0171\pm0.0004$ obtained by the same procedure at the same $D$. The sum rule of eq.~\eqref{eq:sum_rule_Wk}, $W_0+2\sum_{k\ge1}W_k=0$ with $W_0=(D-1)/2$, is satisfied to a relative accuracy of $10^{-16}$, providing an independent check of the construction.

It is worth noting that while the exponent is common to the two models, the amplitude of the $k^{-2}$ law is not: for raw spectra we find $W_k^{\rm SYK}/W_k^{\rm GUE}=1.0652\pm0.0014$ averaged over $k\in[10,100]$, a deviation far outside the statistical uncertainty, whereas after unfolding the same ratio becomes $0.9988$. The prefactor of the diagonal hierarchy therefore retains information about the density of states, and it is unfolding that renders it universal. This is consistent with the logic of section~\ref{sec:the_hypothesis}, where the kernel-universality hypothesis is formulated for unfolded spectra only.

The top-right panel of Fig.~\ref{fig:four_figures_SYK} shows $\log_{10}|M_{ij}|$ for the overlap matrix $M=O_{\rm unfold}O_{\rm uni}^{\sf T}$ of eq.~\eqref{eq:poly_overlap}, constructed from a single unfolded SYK realization. The matrix is sharply localized about the principal diagonal, exactly as for the Gaussian ensembles in Fig.~\ref{fig:M_ij_matrixplot}.

A sharper test is provided by the normalized difference $\delta W_k$ of eq.~\eqref{eq:delta_W}, shown in the bottom-right panel of
Fig.~\ref{fig:four_figures_SYK}. The entire departure of the unfolded SYK kernel from the uniform-lattice kernel is confined to the first few off-diagonals, in close quantitative agreement with the GUE. The approximate kernel universality hypothesis therefore holds for SYK.

\subsection*{Spread complexity}

\begin{figure}
    \centering
    \includegraphics[width=1\linewidth]{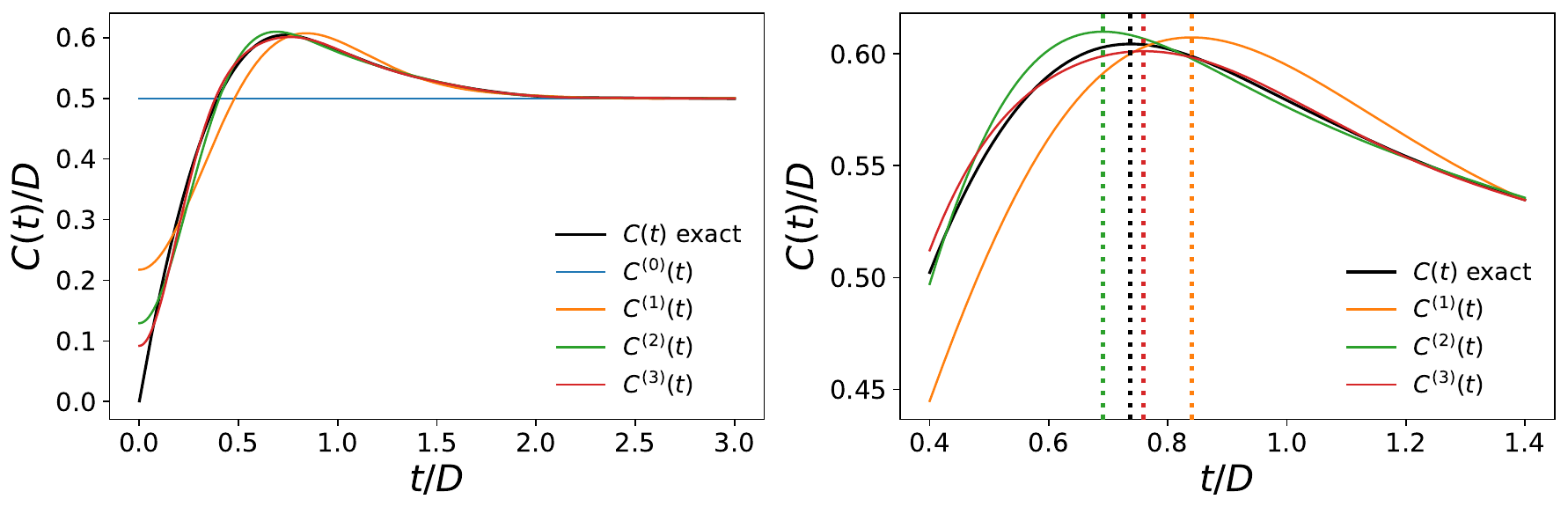}
    \caption{Same as Fig.~\ref{fig:Krylov_comparison}, for the Majorana SYK model with $N=22$ fermions in the positive-parity sector, $D=2^{N/2-1}=1024$, averaged over $50$ disorder realizations. The spectrum is not unfolded; since $J$ only sets an arbitrary energy unit, each realization is rescaled so that its central mean level density matches the GUE normalization of Fig.~\ref{fig:Krylov_comparison}, making $t/D$ directly comparable. The exact complexity is again reproduced by the first few kernel diagonals.}
    \label{fig:syk_complexity}
\end{figure}

Fig.~\ref{fig:syk_complexity} compares the exact spread complexity of the positive-parity SYK spectrum with the truncations $C^{(m)}(t)$, and Table~\ref{tab:syk_peak} lists the corresponding peak positions and heights. As for the Gaussian ensembles, the successive approximations approach the exact result rapidly.

Because the coupling $J$ in eq.~\eqref{eq:syk_variance} sets an arbitrary energy unit, the raw peak position is only meaningful once that unit is fixed; in Fig.~\ref{fig:syk_complexity} each realization is rescaled so that its central mean level density matches the GUE normalization used earlier. The peak height is unaffected by this choice. A convention-free comparison is obtained by working with the unfolded spectrum and the unfolded time $\tau$ of section~\ref{sec:analytic_peak}, for which we find
\begin{equation}
  \tau^{\rm SYK}_{\rm peak} = 2.621 , \qquad
  \frac{C^{\rm SYK}_{\rm peak}}{D} = 0.6291 ,
\end{equation}
against $\tau^{\rm GUE}_{\rm peak}=2.634$ and $C^{\rm GUE}_{\rm peak}/D=0.6286$ obtained numerically at the same $D$. The leading-order analytic prediction of eqs.~\eqref{eq:GUE_peak_time} and~\eqref{eq:GUE_peak_height}, $\tau_{\rm peak}=\sqrt{24/\pi}\simeq2.764$ and $C_{\rm peak}/D=\tfrac12+\tfrac23 e^{-3/2}\simeq0.649$, lies slightly above both, the same small overshoot already visible for the Gaussian ensembles in Fig.~\ref{fig:complexity_GOE_GUE}.

Taken together, these results show that the framework developed in sections~\ref{sec:banded_kernel}--\ref{sec:analytic_peak} applies without modification to a genuine interacting many-body Hamiltonian. The Krylov kernel of the SYK model is banded, its integrated diagonal weights follow the $k^{-2}$ hierarchy, the unfolded kernel is well approximated by that of the uniform lattice, and the resulting complexity peak is quantitatively described by the nearest-neighbour spacing distribution of the appropriate symmetry class.
\begin{table}[t]
\centering
\begin{tabular}{lcc}
\hline
Method & $t_{\rm peak}/D$ & $C(t_{\rm peak})/D$ \\
\hline
Exact    & 0.73748             & 0.60441 \\
Approx 1 & 0.84013 (+13.92\%)   & 0.60735 (+0.49\%) \\
Approx 2 & 0.69146 ($-$6.24\%) & 0.60991 (+0.91\%) \\
Approx 3 & 0.75966 (+3.01\%)   & 0.60119 ($-$0.53\%) \\
Approx 4 & 0.74072 (+0.44\%)   & 0.60608 (+0.28\%) \\
Approx 5 & 0.72251 ($-$2.03\%) & 0.60415 ($-$0.04\%) \\
Approx 6 & 0.74997 (+1.69\%)   & 0.60422 ($-$0.03\%) \\
\hline
\end{tabular}
\caption{Peak position and height of the exact spread complexity of the
positive-parity SYK spectrum ($N=22$, $D=1024$, $50$ disorder realizations),
compared with the successive approximations $C^{(m)}(t)$. The numbers in
parentheses denote the percentage deviations relative to the exact result.}
\label{tab:syk_peak}
\end{table}

\section{Closing remarks}
\label{sec:discussions}
The approximately banded structure of the Krylov kernel suggests several natural directions for future work. First, it would be interesting to investigate whether similar diagonal dominance persists in other models with sparse interactions and tunable randomness like sparse spin models~\cite{Hanada:2023rkf, Hanada:2025pis, Basu:2025ubf}, and to understand how the effective bandwidth depends on the microscopic data. Second, the bandedness may enable analytic bounds on the growth \cite{Parker:2018yvk} of spread complexity by controlling the net weight carried by distant off-diagonals, potentially leading to model-independent constraints on late-time behavior. Third, the detailed pattern of diagonal dominance itself may provide a diagnostic that helps distinguish chaotic from integrable dynamics, complementing standard probes based on level statistics. Fourth, one may view the level index as defining an approximately uniform lattice that can be nonlinearly deformed using the orthogonal-polynomial map; it would be valuable to explore whether this deformation can be exploited to recover the GUE kernel analytically, or to interpolate between different universality classes~\cite{Pandey:1983, Mehta:1983}. Finally, our diagonal decomposition invites generalizations to other notions of complexity defined from spectral data, including weighted or coarse-grained variants that emphasize selected energy scales (temperature) or probe different operator sectors.

From a holographic perspective, it would be interesting to ask whether the quasi-local structure of the energy-space Krylov kernel admits a natural bulk interpretation. In double-scaled SYK and JT gravity, the spread complexity of the thermofield-double state has been identified with the length of the two-sided wormhole, whose finite-dimensional evolution exhibits a maximum
followed by a decrease toward its late-time plateau~\cite{Rabinovici:2023yex, Balasubramanian:2024lqk, Akers:2025ynh}. Our results point to a more microscopic question: can this bulk evolution be decomposed into contributions associated with fixed separations in the ordered boundary spectrum? In particular, it would be valuable to determine whether the approximate $k^{-2}$ hierarchy of integrated kernel weights provides a spectral explanation for the maximal wormhole size and its subsequent relaxation. Establishing such a relation would sharpen the connection between local spectral correlations and the late-time dynamics of the black-hole
interior.

\section*{Acknowledgements}

We thank the participants of the 15th Joburg Workshop (Wits Rural 2025), including the international participants Leopoldo A.~Pando Zayas, Bitan Roy, Juan F. Pedraza, Chethan Krishnan, and Antonio Garc\'\i a-Garc\'\i a, for many motivating discussions related to chaos and complexity. We also thank MITP, CoE-MaSS, and NiTheCS for continued support.  PB also thanks Jyotirmoy Bhattacharya (IIT-KGP) and Subhra Sen Gupta (Shiv Nadar University) for discussions and support, where an initial version of some of these ideas was presented. PB thanks Anurag Sarkar and Sam Van Leuven for discussion. SD is supported by the ‘Atracción de Talento’ program of the Comunidad de Madrid under grant 2020-T1/TIC-20495, the Spanish Agencia Estatal de Investigación through grants CEX2025-001574-S, PID2021-123017NB-I00 and PID2024-156043NB-I00, funded by MCIN/AEI/10.13039/501100011033, and ERDF, EU.

\section*{Data and Repository}

Code related to the energy-space Krylov kernel is available at:
 
\href{https://github.com/pallabbasu/KrylovKernel}{github.com/pallabbasu/KrylovKernel}.

\appendix

\section{Large-\texorpdfstring{$D$}{D} Analysis of the Uniform Kernel}
\label{app:uniform_kernel_sums}

In this appendix we derive the large-$D$ diagonal sums of the uniform-lattice Krylov kernel used in eq.~\eqref{eq:Kernel_sum}. We also show that, the kernel exhibits a semicircular profile.

For convenience, we work with the centered lattice
\begin{equation}
x_i
=
i-\frac{D-1}{2},
\qquad
i=0,\ldots,D-1,
\label{eq:app_centered_lattice}
\end{equation}
which differs from the lattice $\{0,1,\ldots,D-1\}$ only by an overall translation. For odd $D=2L+1$, this becomes the symmetric
lattice $\{-L,\ldots,L\}$. Let $\widehat X$ denote the corresponding position operator,
\begin{equation}
\widehat X|i\rangle
=
x_i|i\rangle.
\label{eq:app_position_operator}
\end{equation}
In the Krylov basis, the off-diagonal recurrence coefficients of $\widehat X$ are
\begin{equation}
\begin{gathered}
\beta_n
=
\frac{n+1}{2}
\sqrt{
\frac{(D-1-n)(D+n+1)}
{(2n+1)(2n+3)}
},
\\
n=0,\ldots,D-2,
\end{gathered}
\label{eq:app_beta}
\end{equation}
with
\begin{equation}
\beta_{-1}
=
\beta_{D-1}
=
0.
\label{eq:app_beta_boundary}
\end{equation}
Using the three-term recurrence relation for the complexity operator $\widehat{\mathcal C}= \sum_{n=0}^{D-1} n\,|K_n\rangle\langle K_n|$, one finds
\begin{equation}
[X,[X,\widehat{\mathcal C}]]
=
2\sum_{n=0}^{D-1}
\left(\beta_{n-1}^{2}-\beta_n^{2}\right)
|K_n\rangle\langle K_n|.
\label{eq:app_double_commutator}
\end{equation}
The difference of the recurrence coefficients is
\begin{equation}
\beta_{n-1}^{2}-\beta_n^{2}
=
\frac{n}{8}
+
\frac{1}{16}
+
\frac{4D^2-1}
{16(2n-1)(2n+1)(2n+3)}.
\label{eq:app_beta_difference}
\end{equation}
Substituting eq.~\eqref{eq:app_beta_difference} into eq.~\eqref{eq:app_double_commutator} gives the exact operator identity
\begin{equation}
4[\widehat X,[\widehat X,\widehat{\mathcal C}]]
-
\widehat{\mathcal C}
=
\frac{1}{2}\,\mathbb I
+
\widehat{\mathcal B},
\label{eq:app_operator_identity}
\end{equation}
where
\begin{equation}
\begin{gathered}
\widehat{\mathcal B}
=
\sum_{n=0}^{D-1}
B_n\,|K_n\rangle\langle K_n|,
\\
B_n
\equiv
\frac{4D^2-1}
{2(2n-1)(2n+1)(2n+3)}.
\end{gathered}
\label{eq:app_B_operator}
\end{equation}
This identity \eref{eq:app_operator_identity} is significant because it relates a divergent sum to a convergent one and isolates the large-$D$ behavior of the left-hand side.

Taking matrix elements of eq.~\eqref{eq:app_operator_identity} in the lattice basis and using
\begin{equation}
x_i-x_j=i-j,
\end{equation}
we obtain
\begin{equation}
\left[4(i-j)^2-1\right]
\langle i|\widehat{\mathcal C}|j\rangle
=
\frac{1}{2}\delta_{ij}
+
\langle i|\widehat{\mathcal B}|j\rangle.
\label{eq:app_kernel_identity}
\end{equation}
Dividing by $D$ and recalling $\mathcal K_{\rm uni}(i,j)\equiv\frac{1}{D}\langle i|\widehat{\mathcal C}|j\rangle$, for $k\geq1$ summing along the $k$th off-diagonal gives the exact relation
\begin{equation}
\left(4k^2-1\right)
\sum_{i=0}^{D-1-k}
\mathcal K_{\rm uni}(i,i+k)
=
\frac{1}{D}
\sum_{i=0}^{D-1-k}
\langle i|\widehat{\mathcal B}|i+k\rangle.
\label{eq:app_diagonal_relation}
\end{equation}
To estimate the right-hand side, we write
\begin{equation}
\sum_{i=0}^{D-1-k}
\langle i|\widehat{\mathcal B}|i+k\rangle
=
\sum_{n=0}^{D-1}
B_n\,S_{n,k}^{(D)},
\label{eq:app_B_shifted_trace}
\end{equation}
with $B_n$ as in eq.~\eqref{eq:app_B_operator}, and
\begin{equation}
S_{n,k}^{(D)}
=
\sum_{i=0}^{D-1-k}
\phi_n(i)\phi_n(i+k).
\label{eq:app_shifted_overlap}
\end{equation}
For fixed $n$ with $k$ fixed, the normalized Gram--polynomial wavefunctions admit a smooth continuum limit on the macroscopic lattice scale. Consequently,
\begin{equation}
S_{n,k}^{(D)}
\longrightarrow
1,
\qquad
D\longrightarrow\infty,
\label{eq:app_shifted_overlap_asymptotic}
\end{equation}
where the limit is taken at fixed $n$ and $k$. Moreover, the coefficients $B_n/D^2$ have a finite large-$D$ limit for fixed $n$
and decay as $n^{-3}$. The leading $O(D^2)$ contribution to eq.~\eqref{eq:app_B_shifted_trace} is therefore controlled by
polynomial degrees that remain finite as $D\to\infty$. It follows that, for fixed $k$,
\begin{equation}
\sum_{i=0}^{D-1-k}
\langle i|\widehat{\mathcal B}|i+k\rangle
=
\operatorname{Tr}\widehat{\mathcal B}
(1+
O(k/D)).
\label{eq:app_shifted_trace_asymptotic}
\end{equation}
Equation~\eqref{eq:app_shifted_trace_asymptotic} is the only large-$D$ asymptotic input used below. All preceding operator
identities, as well as the evaluation of $\operatorname{Tr}\widehat{\mathcal B}$, are exact.

Since $\widehat{\mathcal B}$ is diagonal in the Krylov basis, its trace is
\begin{equation}
\operatorname{Tr}\widehat{\mathcal B}
=
\sum_{n=0}^{D-1}
B_n
=
\frac{4D^2-1}{2}
\sum_{n=0}^{D-1}
\frac{1}
{(2n-1)(2n+1)(2n+3)}.
\label{eq:app_B_trace}
\end{equation}
The remaining sum can be evaluated by defining
\begin{equation}
f(n)
=
\frac{1}{(2n-1)(2n+1)}.
\label{eq:app_f_definition}
\end{equation}
One then has
\begin{equation}
f(n)-f(n+1)
=
\frac{4}
{(2n-1)(2n+1)(2n+3)},
\label{eq:app_telescoping_identity}
\end{equation}
and hence
\begin{align}
\sum_{n=0}^{D-1}
\frac{1}
{(2n-1)(2n+1)(2n+3)}
&=
\frac{1}{4}
\sum_{n=0}^{D-1}
\left[
f(n)-f(n+1)
\right]
\nonumber\\
&=
\frac{1}{4}
\left[
f(0)-f(D)
\right]
\nonumber\\
&=
-\frac{1}{4}
\left(
1+\frac{1}{4D^2-1}
\right)
\nonumber\\
&=
-\frac{D^2}{4D^2-1}.
\label{eq:app_telescoping_sum}
\end{align}
Substituting eq.~\eqref{eq:app_telescoping_sum} into eq.~\eqref{eq:app_B_trace} gives the exact result
\begin{equation}
\operatorname{Tr}\widehat{\mathcal B}
=
-\frac{D^2}{2}.
\label{eq:app_B_trace_result}
\end{equation}
Using eqs.~\eqref{eq:app_shifted_trace_asymptotic} and
\eqref{eq:app_B_trace_result} together with
eq.~\eqref{eq:app_diagonal_relation}, we obtain
\begin{equation}
\begin{gathered}
\sum_{i=0}^{D-1-k}
\mathcal K_{\rm uni}(i,i+k)
\\
=
\frac{1}{D}\left(-\frac{D^2}{2}+o(D^2)\right)
\Big/\left(4k^2-1\right)
\\
=
-\frac{D}{2(4k^2-1)}
+
o(D),
\\
D\longrightarrow\infty,
\end{gathered}
\label{eq:app_offdiagonal_sum}
\end{equation}
where the limit is taken at fixed $k$. Equivalently,
\begin{equation}
\sum_{i=0}^{D-1-k}
\mathcal K_{\rm uni}(i,i+k)
\sim
-\frac{D}{2(4k^2-1)}.
\label{eq:app_offdiagonal_sum_leading}
\end{equation}

The principal-diagonal sum follows directly from the trace of the Krylov position operator:
\begin{align}
\sum_{i=0}^{D-1}
\mathcal K_{\rm uni}(i,i)=
\frac{1}{D}
\operatorname{Tr}\widehat{\mathcal C}=
\frac{1}{D}
\sum_{n=0}^{D-1}n=
\frac{D-1}{2}.
\label{eq:app_diagonal_sum_exact}
\end{align}
Consequently,
\begin{equation}
\sum_{i=0}^{D-1}
\mathcal K_{\rm uni}(i,i)
\sim
\frac{D}{2}.
\label{eq:app_diagonal_sum_largeD}
\end{equation}
Combining eqs.~\eqref{eq:app_offdiagonal_sum_leading} and \eqref{eq:app_diagonal_sum_largeD}, we finally obtain
\begin{equation}
\begin{gathered}
\sum_{i=0}^{D-1}
\mathcal K_{\rm uni}(i,i)
\sim
\frac{D}{2},
\\
\sum_{i=0}^{D-1-k}
\mathcal K_{\rm uni}(i,i+k)
\sim
-\frac{D}{2(4k^2-1)},
\end{gathered}
\label{eq:app_uniform_kernel_sums_final}
\end{equation}
for fixed $k$ in the large-$D$ limit. These are the uniform-kernel diagonal sums used in eq.~\eqref{eq:Kernel_sum}.

Beyond the trace-level asymptotics above, the semicircular profile of the uniform kernel can be exhibited directly on the principal
diagonal by taking the continuum limit at fixed $x\equiv i/D\in(0,1)$. In this limit, the Gram polynomials reduce to shifted Legendre polynomials. Evaluating $\frac{1}{D}\langle i|\widehat{\mathcal B}|i\rangle$ and summing over $n$, we find
\begin{align}
\frac{1}{D} \langle i|\widehat{\mathcal B}|i\rangle
&\simeq
\frac{4D^2-1}{2D^2}
\sum_{n=0}^\infty
\frac{P_n^2(2x-1)}
{(2n-1)(2n+3)}
\nonumber\\
&\xrightarrow[D\to\infty]{}
-\frac{2}{\pi}
\sqrt{1-(2x-1)^2}\, .
\label{eq:semi}
\end{align}
Thus, at fixed $x$, the diagonal matrix elements of $\widehat{\mathcal B}/D$ approach a negative semicircular profile. Using eq.~\eqref{eq:app_kernel_identity}, this implies
\begin{equation}
\mathcal K_{\rm uni}(i,i)
\longrightarrow
\frac{2}{\pi}
\sqrt{1-(2x-1)^2},
\end{equation}
showing that the principal diagonal of the uniform Krylov kernel approaches a semicircular profile in the large-$D$ limit.

\section{Relation between folded and unfolded time}
\label{app:folded_unfolded_time}

The peak positions derived in section~\ref{sec:analytic_peak} are expressed in terms of the dimensionless unfolded time $\tau$, conjugate to the unfolded energies $\{\varepsilon_i\}$ of unit mean spacing. The spread complexity of the original Hamiltonian, however, is computed using the folded energies $\{E_i\}$ and the corresponding physical time $t$. Since unfolding is a nonlinear reparametrization of the spectrum, the relation between these two time variables is in general energy dependent.

Unfolding is defined through the smooth spectral counting function $N(E)$, such that $\varepsilon_i\simeq N(E_i)$ up to an irrelevant additive constant. Since
\begin{equation}
    d\varepsilon=\rho(E)\,dE,
\end{equation}
where $\rho(E)=dN/dE$ is the mean level density, a local energy difference satisfies
\begin{equation}
    \Delta E
    \simeq
    \frac{\Delta\varepsilon}{\rho(E)}.
\end{equation}
Consequently,
\begin{equation}
    (E_{i+k}-E_i)t
    \simeq
    (\varepsilon_{i+k}-\varepsilon_i)
    \frac{t}{\rho(E_i)}.
    \label{eq:local-rescaling}
\end{equation}
Thus, locally, the unfolded time is related to the physical time by $\tau\simeq t/\rho(E)$. For the GOE/GUE normalization used in this work, the mean density is
\begin{equation}
    \rho(E)
    =
    \frac{D}{2\pi}\sqrt{4-E^2},
    \qquad
    E\in[-2,2],
\end{equation}
and hence
\begin{equation}
    \rho(0)=\frac{D}{\pi}.
    \label{eq:rho-zero}
\end{equation}
Approximating the local density by its bulk value therefore gives
\begin{equation}
    \tau
    \simeq
    \frac{\pi}{D}\,t,
    \qquad\text{or equivalently}\qquad
    \frac{t}{D}
    \simeq
    \frac{\tau}{\pi}.
    \label{eq:tau-t-relation}
\end{equation}
Applying eq.~\eqref{eq:tau-t-relation} to the leading-order peak positions $\tau_{\rm peak}^{\rm GUE}\simeq2.764$ and $\tau_{\rm peak}^{\rm GOE}\simeq2.662$ gives
\begin{equation}
    \frac{t_{\rm peak}^{\rm GUE}}{D}
    \simeq 0.880,
    \qquad
    \frac{t_{\rm peak}^{\rm GOE}}{D}
    \simeq 0.847.
\end{equation}
These values are close to the peak position obtained from the nearest-neighbour approximation in table~\ref{tab:peak_comparison}.
\begin{figure}[h]
    \centering
    \includegraphics[width=0.65\linewidth]
    {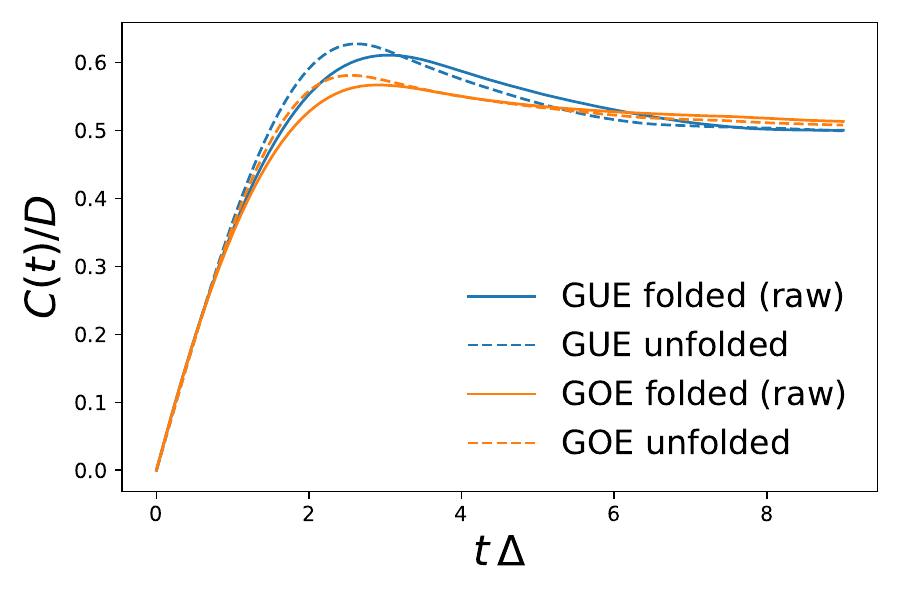}
    \caption{Comparison of the full spread complexity computed from the
    original (folded) and unfolded spectra for the GOE and GUE ensembles
    with $D=1000$, averaged over $20$ disorder realizations. Time is shown
    in units of the spectrum-averaged level spacing~\eqref{eq:delta-def},
    with $\Delta=4/(D-1)$ for the folded spectra and $\Delta=1$ for the
    unfolded spectra.}
    \label{fig:folded-unfolded}
\end{figure}

For the numerical comparison of folded and unfolded spectra, it is convenient to introduce instead the dimensionless variable $t\Delta$, where
\begin{equation}
    \Delta
    \equiv
    \frac{E_{\max}-E_{\min}}{D-1}
    \label{eq:delta-def}
\end{equation}
is the spectrum-averaged level spacing. For the present normalization,
\begin{equation}
    \Delta_{\rm folded}
    =
    \frac{4}{D-1},
    \qquad
    \Delta_{\rm unfolded}
    =
    1.
    \label{eq:delta-values}
\end{equation}

Fig.~\ref{fig:folded-unfolded} compares the corresponding complexities as functions of $t\Delta$. Although both display the same growth--peak--plateau structure, with comparable peak heights and positions, they are not equivalent, and this constitutes an important limitation in interpreting the unfolded results. Unfolding maps $H$ to $H_{\rm unfold}=N(H)$, which is in general a nonlinear function of $H$; it therefore not only rescales energy
differences by the local density $\rho(E)$ (see \eref{eq:local-rescaling}),
but also changes the Lanczos polynomials and hence the Krylov basis. The two
complexities thus measure spreading along related but distinct Krylov chains,
and cannot in general be related by a rescaling of time. Only for an affine
map $E \to aE + b$ with $a > 0$ do the two Krylov bases coincide, in which
case the complexities differ only by a rescaling of time.

Consequently, the analytic results of Sec.~\ref{sec:analytic_peak} describe
the spread complexity of $H_{\rm unfold}$, not that of $H$ itself, and apply
to the original Hamiltonian only qualitatively. This choice is deliberate: the
spread complexity of the raw spectrum depends both on the level correlations
and on the smooth density of states, and the latter alone can produce a peak
even in the absence of random-matrix correlations, as shown for a logarithmic
spectrum and for saddle-dominated integrable models
in~\cite{Erdmenger:2026iga}. Unfolding removes this non-universal
contribution and isolates the dependence on the level statistics, which is
the subject of this work.

However, the folded Krylov kernel is itself banded
(section~\ref{sec:banded_kernel}), so that only nearby levels contribute
significantly. The folded complexity may therefore be approximated by a
superposition of the unfolded one over local time rescalings $t/\rho(E)$. A
study of this relation, and of approximate kernel universality for folded
spectra, is left for future work.

\section{Poisson spectra}
\label{app:poisson_checks}

Here we repeat the tests of Sec.~\ref{sec:the_hypothesis} for Poisson spectra. Unlike the Gaussian ensembles, Poisson spectra lack spectral rigidity: the number variance $\Sigma^2(L)$, i.e., the variance of the number of unfolded levels in an interval of length $L$, grows linearly, $\Sigma^2(L) \simeq L$, whereas for the Gaussian ensembles it grows only logarithmically. Consequently, the typical deviations $\delta_i$ in eq.~\eqref{eq:unfolded_near_lattice} grow as $\sqrt{D}$, rather than as $\sqrt{\ln D}$. We draw $D = 1000$ independent levels from the same semicircle density as the GUE spectra of the main text, so that only the level correlations differ, and unfold them with the same procedure.
\begin{figure}
    \centering
    \includegraphics[width=0.7\linewidth]{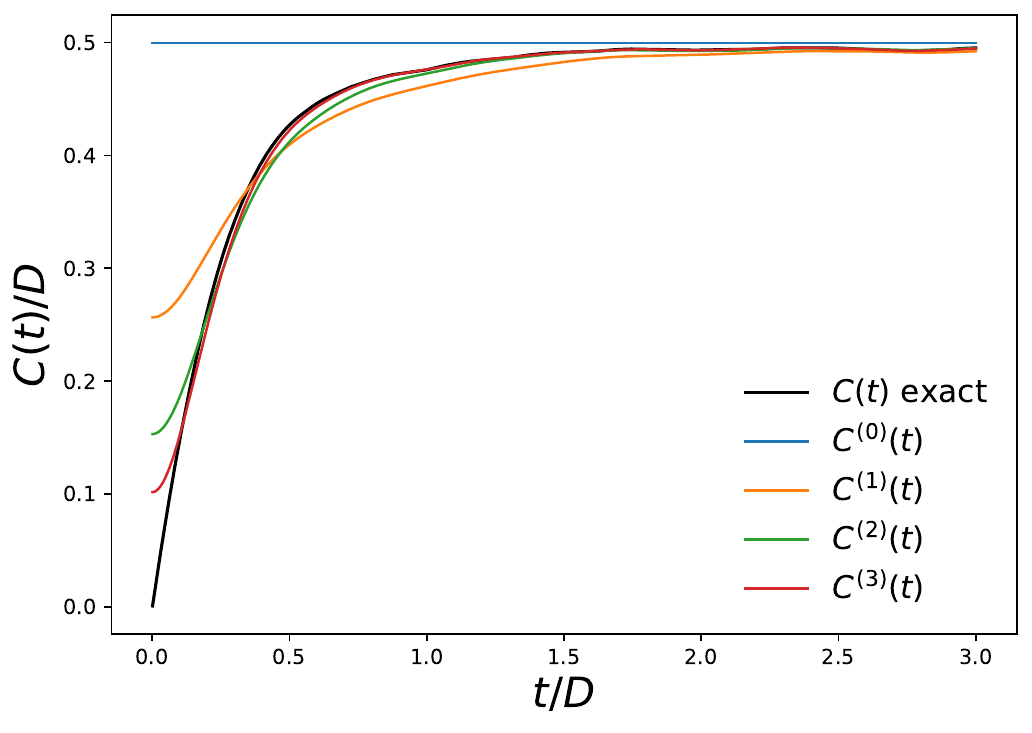}
    \caption{Same as the left panel of Fig.~\ref{fig:Krylov_comparison}, for Poisson spectra (not unfolded), averaged over 20 realizations ($D = 1000$).}
    \label{fig:poisson_convergence}
\end{figure}

\begin{figure}
    \centering
    \includegraphics[width=1\linewidth]{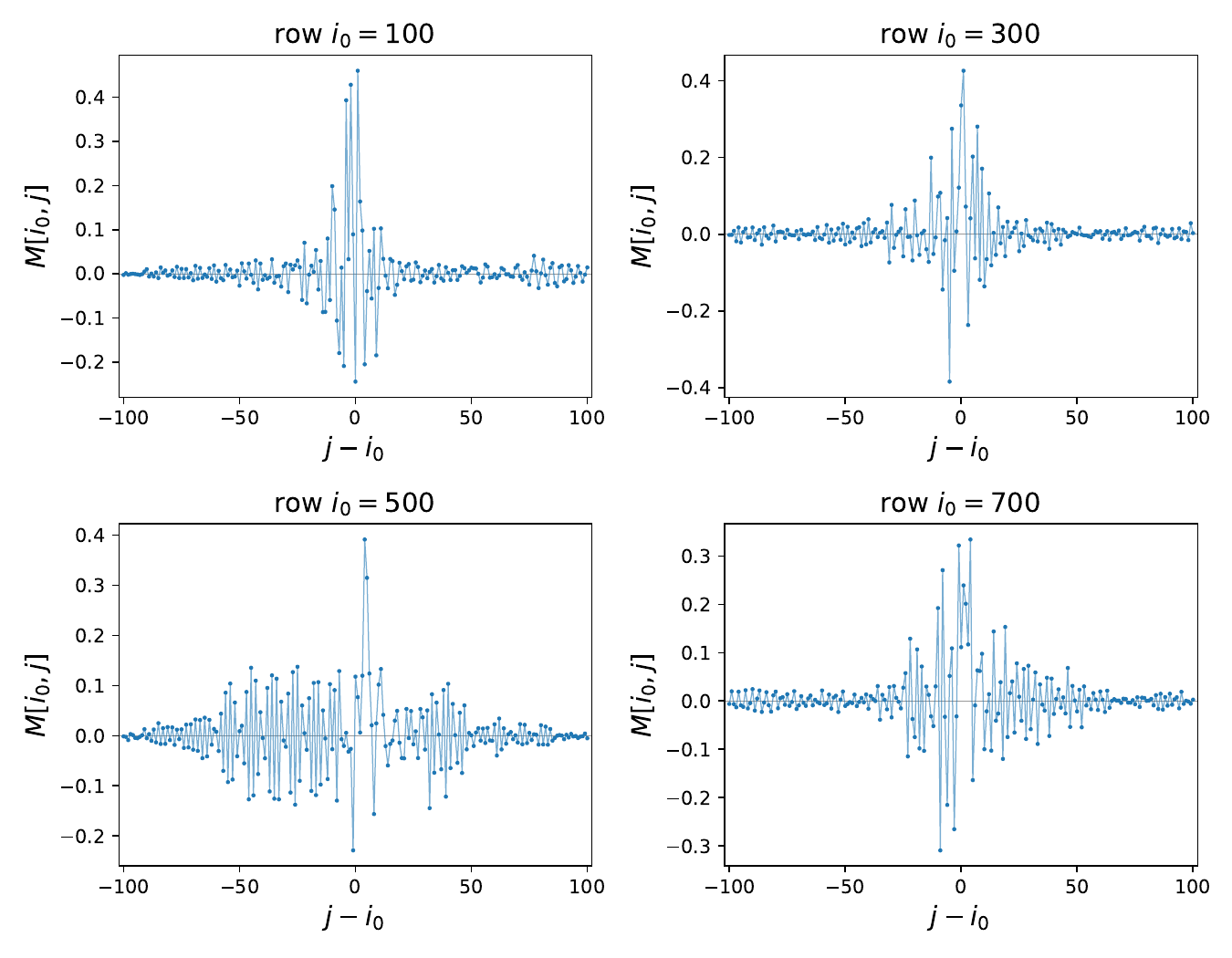}
    \caption{Same as Fig.~\ref{fig:M_different_diagonal}, for a single unfolded Poisson realization. The rows are broader than for the GUE, and their maxima are displaced from $j = i_0$ by a few sites.}
    \label{fig:poisson_M_rows}
\end{figure}

\begin{figure}
    \centering
    \includegraphics[width=1\linewidth]{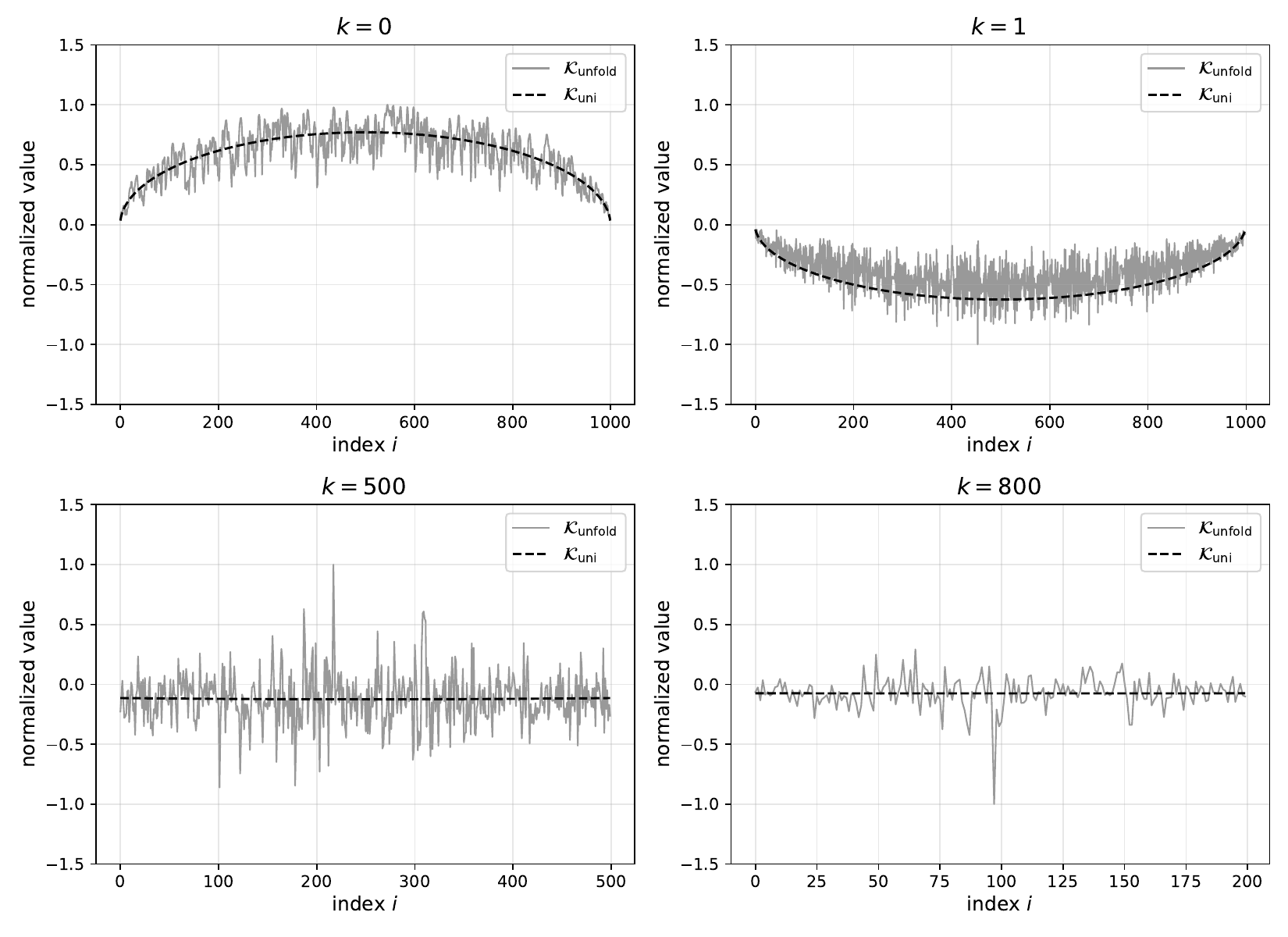}
    \caption{Same as Fig.~\ref{fig:comparison_Kernel}, for the Poisson realization of Fig.~\ref{fig:poisson_M_rows}.}
    \label{fig:poisson_kernel_profiles}
\end{figure}

Fig.~\ref{fig:poisson_convergence} shows that the diagonal expansion, eq.~\eqref{eq:complexity_sum1}, converges rapidly, as for the GUE, while the complexity approaches its plateau without developing a peak, in agreement with Sec.~\ref{sec:Poisson_no_peak}. The overlap matrix $M$ remains concentrated around the principal diagonal (Fig.~\ref{fig:poisson_M_rows}), but its rows are broader than for the GUE, and their maxima are displaced by a few sites, following the deviations $\delta_i$; moreover, this localization does not improve with increasing $D$. The uniform-lattice kernel accurately follows the semicircular envelope of the unfolded kernel on the principal diagonal and the mean of its fluctuations for large $k$ (Fig.~\ref{fig:poisson_kernel_profiles}). On the first off-diagonal, however, it overestimates the magnitude of the unfolded kernel, corresponding to $\delta W_1 < 0$; as Fig.~\ref{fig:deltaW_k} shows, such discrepancies extend over the first few diagonals.

The hypothesis therefore holds for Poisson spectra in a weaker sense than for the Gaussian ensembles: the overlap matrix is less localized, and the discrepancy $\delta W_k$ is larger. Nevertheless, $\delta W_k$ remains confined to the first few diagonals and depends only weakly on $D$, and the uniform-lattice kernel reproduces the qualitative behavior of the Poisson spread complexity, namely its monotonic growth toward the plateau without a peak, with small quantitative deviations in the growth regime (Fig.~\ref{fig:folded_unfolded_poisson}).

\bibliography{bibliography}

\end{document}